\documentstyle[12pt,draft,epsf]{article}

\newcommand \be  {\begin{equation}}
\newcommand \ee  {\end{equation}}
\newcommand \bea {\begin{eqnarray} \nonumber }
\newcommand \eea {\end{eqnarray}}

\newcommand \ga  {\alpha}
\newcommand \gb  {\beta}
\newcommand \gd  {\delta}

\newcommand \get {\eta}

\newcommand \gl  {\lambda}

\newcommand \gr  {\rho}
\newcommand \gs  {\sigma}
\newcommand \gth {\theta}
\newcommand \gD  {\Delta}
\newcommand \gG  {\Gamma}
\newcommand \gL  {\Lambda}
\newcommand \gO  {\Omega}

\newcommand \cC  {{\cal C}}

\newcommand \cL  {{\cal L}}
\newcommand \cN  {{\cal N}}
\newcommand \cP  {{\cal P}}
\newcommand \cR  {{\cal R}}
\newcommand \cS  {{\cal S}}
\newcommand \cX  {{\cal X}}

\newcommand \NE {\not=}
\newcommand \Tr {\mbox{Tr}}

\newcommand \lan {\langle}
\newcommand \ran {\rangle}
\newcommand \bra {\langle 0 |}
\newcommand \ket {| 0 \rangle }

\newcommand{\AN}{A^{(N)}}
\newcommand{\zN}{z_c^{(N)}}

\begin{document}

\title{Replica Theory and Large $D$ Josephson Junction
Hypercubic Models}
\author{ Enzo Marinari$^{(C)}$, Giorgio Parisi$^{(R)}$
and Felix Ritort$^{(M)}$ \\
{\small $(C)$: Dipartimento di Fisica and Infn, Universit\`a di Cagliari}\\
{\small Via Ospedale 72, 09100 Cagliari (Italy)}\\
{\small \tt marinari@ca.infn.it}\\
{\small $(R)$: Dipartimento di Fisica and Infn,
Universit\`a di Roma {\em La Sapienza}}\\
{\small Piazzale Aldo Moro 2, 00185 Roma (Italy)}\\
{\small \tt parisi@roma1.infn.it}\\
{\small $(M)$: Departamento de Matematica Aplicada}\\
{\small  Universidad Carlos III de Madrid}\\
{\small  Calle Butarque 15, 28911 Leganes (Madrid) (Spain)}\\
{\small \tt ritort@dulcinea.uc3m.es}
}
\date{February 4, 1995}
\maketitle

\begin{abstract}
We study the statistical mechanics of a $D$-dimensional array of Josephson
junctions in presence of a magnetic field on a lattice of side $2$. In the
high temperature region the thermodynamical properties can be computed in the
limit $D \to \infty$. A conjectural form of the thermodynamic properties in
the low temperature phase is obtained by assuming that they are the same of
an appropriate spin glass system, based on quenched disordered couplings.
Numerical simulations show that this conjecture is very accurate in one regime
of the magnetic field, while it is probably slightly inaccurate in a second
regime.
\end{abstract}

\vfill

\begin{flushright}
  { \tt cond-mat 9502067 }
\end{flushright}
\newpage

\section{\protect\label{S_INT} Introduction}

In this paper we pursue our research program on the relation  among systems
based on a Hamiltonian containing quenched disorder and systems with a fixed
frustrated but non-random  Hamiltonian \cite{MAPARI_1,MAPARI_2}.  Here we
study the statistical mechanics of arrays of Josephson   junctions
\cite{PARISI} in $D$-dimensions, in the limit where $D \to \infty$ on a
lattice of  size $2$ (i.e. on a single hypercube with $2^D$ points).  The case
of a fully  frustrated lattice has been already discussed in reference
\cite{MAPARI_FF}.

In the framework of the  spherical approximation the thermodynamic properties
can be  computed by using the results obtained in \cite{PARISI}.  It is
possible to prove that the spherical approximation gives the correct results
even for the XY model (the one that will mainly interest us here) in the  high
temperature phase. At $T=T_c$ the system undergoes a phase transition. In the
low temperature region the spherical approximation breaks down. We  conjecture
that the thermodynamical properties of the system are the same of an
appropriate spin glass model, constructed in such a way to have the  same high
temperature expansion than the original deterministic model. We solve the
disordered model by using the  replica approach.

We have simulated numerically systems in dimension $D$, ranging from $3$ to
$16$. We find that the comparison of the  numerical simulation with the
theoretical results is extremely good in the high temperature phase (as
expected). In the low temperature phase things seem to work quite well when we
move toward the fully frustrated model (starting from a magnetic field
$\gth=\frac{\pi}{2}$ and increasing $\gth$), but when decreasing $\gth$
(towards the ferromagnetic system)  we find  a rather disturbing phenomenon.
Indeed in this region a naive extrapolation for $D\to\infty$ gives a result
which  differs slightly from the analytic results (obtained by applying
replica theory to the model which contains quenched disorder).  Such
discrepancy becomes  larger and larger when decreasing the frustration.  We
are unable  to decide if in this regime our analytic results are only a good
approximation to the behavior of the system without quenched disorder, or if
they are  exact and the finite $D$ corrections have a peculiar dependence on
$D$. An  analytic computation inside our theoretical framework of the $1 \over
D$  corrections would be extremely useful, but it goes beyond the aims of this
note.

In section (\ref{S_REV}) we give a short summary of the results obtained in
\cite{PARISI}. In  section (\ref{S_STR}) we describe our strategy, and define
the model with quenched disorder which we will {\em substitute} to the original
deterministic model. In  section (\ref{S_HIG}) we will discuss the high $T$
expansion. In  section (\ref{S_LOW}) we will use replica theory to solve the
random model for $T<T_c$. In section (\ref{S_COM}) we will describe our
numerical simulations, and compare them to the analytic results obtained in
the former sections. Finally in the appendix we close a gap in the proof of
ref.~\cite{PARISI} about the connection of the high temperature expansion and
the Green functions of the $q$-deformed harmonic oscillator.

\section{\protect\label{S_REV}
Diagram Counting, Josephson Junctions and $q$-Deformations}

We will start here by defining the relevant statistical models, and by
reviewing in a very cursory manner the results of \cite{PARISI}. The prototype
model is the Gaussian model, defined by the Hamiltonian

\be
  \protect\label{E_GAUHAM}
  \gb H_G \equiv -\gb \Re \{c(D) \sum_{j,k}\eta^*_j U_{j,k} \eta_k\}
  + \frac12 \sum_k |\eta_k|^2\ .
\ee
Here $c(D)$ is a normalization constant, which will be useful later  to
rescale the Hamiltonian in order to obtain a non trivial limit when $D$ goes
to infinity. $c(D)$ will be $\frac1{2D}$ for the usual ferromagnetic
XY model (and in this case we will get a phase transition at $\gb=1$). For a
model with random couplings, and for the frustrated models we will be mainly
discussing in this note,  we will have to take $c(D)\simeq (2D)^{-\frac12}$
in order to insure a sensible infinite dimensional
limit .

Real and imaginary part of the complex $\eta_j$ lattice variables can take
values that range from $-\infty$ to $+\infty$. We will consistently indicate
with $\get$ the fields of the Gaussian model. With $\phi_i$ we will denote the
fields of the XY model, which are constrained to be, on every site, of modulus
$1$, i.e. for all sites $i$

\be
  \protect\label{E_XYCON}
  \ |\phi_i|^2 = 1\ .
\ee
Their dynamics is governed by the Hamiltonian

\be
  \protect\label{E_XYHAM}
  \gb H_{XY} \equiv -\gb \Re \{c(D) \sum_{j,k}\phi^*_j U_{j,k} \phi_k\}\ .
\ee
With $\gs_i$ we will denote the fields of the spherical model,
which satisfy the constraint

\be
  \protect\label{E_SPHCON}
  \sum_i|\gs_i|^2 = N\ ,
\ee
with the Hamiltonian

\be
  \protect\label{E_SPHHAM}
  \gb H_{S} \equiv -\gb \Re \{c(D) \sum_{j,k}\gs^*_j U_{j,k} \gs_k\}\ .
\ee
We can rewrite the spherical model Hamiltonian by including the constraint by
means of a Lagrange multiplier $\mu$. We can write

\be
  \protect\label{E_SPHHCO}
  \gb H_{S} \equiv -\gb \Re \{c(D) \sum_{j,k}\gs^*_j U_{j,k} \gs_k\}
  + \mu (\sum_i|\gs_i|^2 - N)\ ,
\ee
for unconstrained variables. Integration over $\mu$ insures that the spherical
constraint is implemented.

The $U$ couplings are non-zero only for first neighboring site couples. They
are complex numbers of modulus $1$. In the following we will always have that

\be
  \protect\label{E_UBACK}
  U_{j,i}=U^*_{i,j}\ ,
\ee
i.e. the link couplings are oriented, and when coming back on a link one takes
the opposite phase of when following it in the positive direction. By using the
language of gauge theories one says that the $U$ couplings are $U(1)$ lattice
gauge fields \cite{PARISI_DUE}.

We will be discussing here hypercubic models. For a $D$ dimensional model the
field variables live on a $D$-dimensional hypercube, which is done of $2^D$
points. We only include link couplings which are internal to the cube, i.e. we
use open boundary conditions. The number of independent link couplings in our
lattice is $D 2^{D-1}$. The limit $D\to\infty$ is taken by letting the
dimensionality of the hypercube to increase.

Apart from the two cases we already quoted (i.e. the ferromagnetic model with
all $U$ fields equal to $1$ and the XY spin glass, where
$U_{j,k}=\exp(i r_{j,k})$, and the $r_{j,k}$ are random numbers uniformly
distributed in the interval $(0,2\pi]$) we will mainly be interested here in
models where the couplings are such to generate a constant magnetic field
$\gth$. The magnetic field which flows to a given elementary plaquette $\cP$
is\footnote{The plaquette is the elementary lattice closed circuit, done from
$4$ oriented links forming a minimal square.}

\be
  \protect\label{E_PLA}
  \prod_\cP U_\cP = e^{\pm i \gth_\cP}\ ,
\ee
where the sign of the exponents determine if the field is flowing in the
positive or in the negative direction. We will be interested in the case
where $\gth_\cP=\gth$ is constant on all the lattice, i.e. the plaquettes
undergo a constant, uniform frustration. The case $\gth=0$ gives the
ferromagnetic model, while the case $\gth=\pi$ gives the fully frustrated
model, which we have discussed in detail in ref. \cite{MAPARI_FF}. If we let
$\gth_\cP$ to be a random variable we obtain a so-called {\em gauge glass}
\cite{HUSSEU,RTYF,GINGRA}.

The values of the signs of the exponents that enter eq. (\ref{E_PLA}) are
in part arbitrary. Parallel plaquettes have to be cut by a flux flowing in
the same direction, i.e. the signs must have the structure
$\cS_{\alpha,\beta}$, where $\cS$ is a tensor, which is automatically
antisymmetric because of the way we have used to define the $U$ fields.
We are interested in the choice of a {\em generic} structure of $\cS$ (for
the reasons we have discussed in \cite{MAPARI_1,MAPARI_2} and we will discuss
better in the following). We need a generic representative of the ensemble of
the possible choices of $\cS$. One can see that for $D>3$ the choice
$\cS_{\alpha,\beta}=1$ is not a good choice (this is not true in $3$ and $2$
$D$, where all choices of $\cS$ are equivalent). We also need to define the
parameter

\be
  \protect\label{E_Q}
  q \equiv \cos(\gth) \ ,
\ee
which will play an important role in the following.

Let us be more explicit and summarize. Our model lives in a magnetic field
given by the antisymmetric tensor

\be
  \gth_{\ga,\gb} = \cS_{\ga,\gb} \ \gth \ ,
\ee
where in the continuum $\gth_{\ga,\gb}$ becomes  $\partial_\ga A_\gb -
\partial_\gb A_\ga$. This is a condition of complex frustration on the
elementary plaquettes. For $\theta=\pi$ we recover the  fully frustrated model.
On our hypercubic lattice the construction of the $U$ fields that generate a
$\theta$ frustration is unique modulo gauge transformations, and can be easily
given. We define $U_\mu(j)$ the coupling $U$ which goes from the site $j$ in
the direction $\mu$ (we only have first neighbor non-zero coupling). $\mu$ goes
from $1$ to $D$, since we only need to set couplings in the positive direction
(the one going in the negative direction are set by the relation
(\ref{E_UBACK})). We set

\be
  U_1(j)=1\ ,
\ee
and for $\mu>1$

\be
  \protect\label{E_STA}
  U_\mu(j) = e^{i\gth \sum_{\nu=1}^{\mu-1}\cS_{\mu,\nu}j_\nu}\ .
\ee
For example in $4$ dimensions we get

\bea
            U_1(j) &=& 1\ ,\\
  \nonumber U_2(j) &=& e^{i\gth(\cS_{2,1}j_1)}\ ,\\
  \nonumber U_3(j) &=& e^{i\gth(\cS_{3,1}j_1+\cS_{3,2}j_2)}\ ,\\
  \nonumber U_4(j) &=& e^{i\gth(\cS_{4,1}j_1+\cS_{3,2}j_2+\cS_{3,3}j_3)}\ .
\eea
In this note we will obtain a generic $\cS$ by picking up at random the
$\pm1$ components of the antisymmetric tensor $\cS_{\ga,\gb}$. This is only a
small amount of randomness. The system is determined by $D\ 2^{D-1}$ couplings,
i.e. a number of couplings exponentially large in $D$, while we are using
only order of $D^2$ random numbers to pick up phases which make generic the
magnetic field tensor. It is maybe possible to imagine simple forms of the
tensor $\cS$ which give a generic magnetic field (i.e. with the correct
moments).

Let us start from the discussion of the gaussian model, where the $\eta$
fields are unconstrained, (\ref{E_GAUHAM}), and summarize the steps taken in
ref. \cite{PARISI}. We will later introduce the modifications needed to
discuss the XY (\ref{E_XYCON},\ref{E_XYHAM})  and the spherical model
(\ref{E_SPHCON},\ref{E_SPHHAM}). We will assume in the following we are
taking the $D\to\infty$ limit by the hypercubic lattice approach we have
described before.

On general grounds the free energy $F$ of a statistical model like the ones we
have defined in (\ref{E_GAUHAM},\ref{E_XYHAM},\ref{E_SPHHAM}) can be written
through its {\em high temperature expansion} as \cite{PARISI_SFT}

\be
  \gb F(\gb) = \sum_n \frac{(\gb c(D))^n}{n}\  \cN(n)\  \lan W(C) \ran_n\ ,
\ee
where the sum runs over all circuit lengths $n$, $\cN(n)$ is the number of
rooted closed circuits of length $n$, and $\lan W(C) \ran_n$ is the
average over all circuits of length $n$ of the value of Wilson loop $W(C)$
(defined as the oriented product of the couplings that one encounters when
following the closed circuit). We will be interested in the $D\to\infty$
limit, and define

\be
  \protect\label{E_DEFG}
  G^{(q)}_{n} = \lim_{D\to\infty} (2D)^{-\frac{n}{2}} \cN(n)\
  \lan W(C) \ran_n\ .
\ee
We are indicating with the superscript $(q)$ the dependence of $G$ over the
value of $\gth$, i.e. of $q$.
Using this definition in the $D\to\infty$ limit the free energy reads

\be
  \protect\label{E_DEFF}
  \gb F(\gb) = \sum_n \frac{\gb^n G^{(q)}_{n}}{n}\ .
\ee
In order to obtain the free energy of the system we will have to compute the
functions $G^{(q)}_{n}$.

In the ferromagnetic case (where $\gth=0$ and $q=1$) everything is easy, since
$\lan W(C) \ran_n=1$ for all values of $n$. Here it is easy to recover all
the usual results of the high $T$ expansion in the $D\to\infty$ limit
\cite{PARISI}.

The next step can be started by discussing the $D\to\infty$ limit of a
gaussian, XY or spherical spin glass, i.e. the situation where
$U_{j,k}=\exp(i r_{j,k})$, and the $r_{j,k}$ are random numbers uniformly
distributed in the interval $(0,2\pi]$, and one eventually averages over the
$r$ random variables. We have already reminded the reader that this is an
usual spin glass (the replica symmetric solution for the XY case is discussed
already in \cite{SK2}). This is not one of the non-random models that we
want to study here, but we are using it just in order to go back soon to our
models in magnetic field $\gth$ with a bit more knowledge.
In this case one can easily see that
the only (closed) diagrams contributing to the free energy are {\em
backtracking} diagrams. For any steps going to $i$ to $j$ we need the
opposite step going from $j$ to $i$, or the integrals over the quenched $r$
variables gives us zero.

This step is completed by noticing that the backtracking diagrams are
also the only ones which survive (in the $D\to\infty$ limit) in the
$\gth=\frac{\pi}{2}$ model, which we call {\em half frustrated}. Here
on all elementary plaquettes the product of the plaquette couplings is
purely imaginary, $\pm i$. It is easy to see why. In the $D\to\infty$ limit
each step is taken in a different direction. So each time we find a phase $i$
which enters our Wilson loop, we will have to consider the contribution of an
other path with the conjugate phase $-i$ (in $D$ finite two steps in the same
direction can create a situation where this cancellation does not hold
anymore).

In this way we have associated backtracking diagrams to one particular case
of our frustrated models, the one in which $\gth=\frac{\pi}{2}$, and the
plaquette frustration has the constant imaginary value $i$
(apart from a sign). The next step consists of associating to each
backtracking diagram a planar diagram.

The instructions are the following. In order to compute $G^{(0)}_{2n}$ consider
$2n$ letters, equal at couples, i.e. take two $a$, two $b$, two $c$, up to $n$
couples. Form a word by ordering these letters, and put the ordered letters  on
a circle. Now connect equal letters with lines. Count the number of
intersections of these lines. Call $I_n(m)$ the number of words done of $n$
couples which have $m$ intersections. $I_n(m)$ is a topological invariant, and
depends only on the order of the letters. The condition of zero intersections
implies that the diagram is planar. One has that, for the Gaussian model,

\be
  G^{(0)}_{n} = I_n(0)\ .
\ee
This shows \cite{PARISI}
that the problem of the gaussian half-frustrated model
(and of the Gaussian spin glass) is solved by counting planar diagrams.
$I_n(0)$ has been computed in \cite{BIPZ}, and the generalization to the XY
and spherical model is straightforward.

The next step of the deduction of ref. \cite{PARISI} is the one that concerns
our model which lives in a constant magnetic field. It is a generalization of
the counting argument discussed before, and it says that we can solve our
problem by counting non-planar diagrams, i.e. by counting words which have a
non-zero number of intersections. There are two crucial results. The first
states that, in a large number of dimensions $D$,

\be
  G^{(q)}_{2n} = \sum_{w(2n)} q^{A(w(2n))}\ ,
\ee
where $A(w)$ is the signed area associated to the diagram represented by the
word $w$. Planar diagrams, with zero intersections, have zero area. In the
$D\to\infty$ limit all the steps which form the diagram are taken in different
directions, and the projected signed area over the plane $(\mu,\nu)$
$A_{\mu,\nu}$ can only take the values $0$ and $\pm1$. The total area
$A$ has been defined as the sum of the modulus of the individual
signed areas

\be
  A \equiv \sum |A_{\mu,\nu}|\ .
\ee
The second part of this step shows that $A(w)$ is equal to the number of
intersections of the line drawing associated to the word $w$. Considering
diagrams which have a non-zero area means considering words which line drawing
has a non-zero number of intersections.

This generalization of the counting of planar diagrams to a counting on
non-planar diagrams has shown in a last step to have an underlying powerful
algebraic structure. Indeed ref. \cite{PARISI} shows that (and we complete
here the proof of this statement)

\be
  \protect\label{E_GA}
  G^{(q)}_{2n} = \sum_{w(2n)} q^{A(w(2n))} =
  \lan 0 | \  \cX_q^{2n}  | 0 \ran\ ,
\ee
where the operator $\cX_q$ is

\be
  \protect\label{E_X}
  \cX_q = \cR_q + \cL_q \,
\ee
and the operators $\cL_q$ and $\cR_q$ satisfy the commutation relations of the
annihilation and the creation operators of a $q$-deformed harmonic oscillator

\be
  \protect\label{E_COMM}
  \cL_q \cR_q - q \cR_q \cL_q = 1 \ .
\ee
The vacuum state $|0\ran$ (for the model with charge $q$) is defined by the
condition

\be
  \cL_q|0\ran=0\ .
\ee
$\cL_q$ may be identified with the annihilation
operator and $R_q$ with the creation operator for a $q$-deformed harmonic
oscillator. They can be  represented as:

\bea
  \cR_q |m \ran &=&   [m]_q^{\frac{1}{2}} |m +1\ran\ , \\
  \cL_q |m \ran &=& [m-1]_q^{\frac{1}{2}} |m -1\ran\ ,
\eea
where

\be
  [m]_q \equiv \frac{1-q^{m+1}}{1-q}\ ,
\ee
and  $m$ takes integer values in the interval $(0 -\infty]$.

For $q=1$ we have the usual ferromagnet, for $q=-1$ the fully frustrated model
\cite{MAPARI_FF}, and for $q=0$ our half frustrated model (which has the same
diagrammatic expansion than the spin glass model).

These are the basis on which we will try to build here, mainly trying to
gather information about the behavior of these frustrated models in the low
$T$, glassy phase.

\section {\protect\label{S_STR} Our Strategy and the Definition of the
Random Model}

We will use here a strategy we have introduced in \cite{MAPARI_1},
\cite{MAPARI_2} (see also \cite{BOUMEZ} for the development of very
connected ideas). We start with a model which does not contain
quenched disorder, but that is complex enough to make us suspicious of
the possible presence of a spin glass like phase for temperatures $T$
low enough. We look for a model which contains quenched disorder, and
that is similar enough to the original model to have potentially the
same behavior (even in the low $T$ phase, if we are very
ambitious). Replica theory allows us to solve the random model, and to
try and get information about the deterministic model. References
\cite{MAPARI_1} and \cite{MAPARI_2} discuss successful examples of
the use of this strategy.

Here we will adopt the same approach. We will introduce a model
containing random quenched disorder. In this new model the new
$\widehat{U}$ couplings will be chosen at random (as opposed to the
original $U$ couplings which are determined by the deterministic
algorithm (\ref{E_STA}) such to give us the needed complex
frustration). The random values of the $\widehat{U}$ will be selected,
following \cite{MAPARI_2}, such that the new free energy will have the
same high temperature expansion than the original model.  So, we will
be in the typical situation described in \cite{MAPARI_1} and
\cite{MAPARI_2}. We will have a model where the couplings
$\widehat{U}$ will be distributed according to a probability
distribution, determined from the request of finding the same high $T$
expansion than in the original frustrated model. The original model
will be in this way by construction a given (hopefully typical)
realization of the coupling constants constructed according to this
probability distribution.

Because of these remarks, and of our constructive procedure, the
deterministic model and the random one coincide in the high  $T$ phase. We
hope to learn as much as possible about the low $T$ phase, and that the two
models are also in this phase very similar.

We will have to start by computing the high temperature expansion for our
model with complex frustration. Knowing that we will use a  reverse
engineering procedure in order to find out the probability  distribution of
random couplings $\widehat{U}$ that have the same high temperature
expansion.  Finally we will use the replica theory to compute the low
temperature behavior of the random model.  For sake of simplicity we will
present here the computation done under the hypothesis of no replica
symmetry breaking. We will compare these analytic results to numerical
simulations of the frustrated model.

We will consider a model containing quenched disorder that has the same
form of the original model with complex deterministic frustration.  In
the random model the couplings $\widehat{U}$ will be taken randomly
among all matrices having the same spectral distribution of the
deterministic model.  More precisely for finite $D$ we extract a set
of $2^D$ values of the eigenvalues $\gl$, such that

\be
\protect\label{E_RANA}
2^{-D} \sum_{j=1,2^D} \gl_j^n \simeq \int d\gl\  \gr_\gD (\gl)\  \gl^n\ ,
\ee
where $\gr_\gD$ is the spectral density of the Laplacian operator,
and will be discussed in more detail in next section. We finally set

\be
\protect\label{E_RANB}
\widehat{U}_{i,k}=\sum_{j=1,2^D} V^*_{i,j}\  \gl_{j}\  V_{j,k}\ ,
\ee
where $V$ is a random unitary matrix i a $2^D$ dimensional space.

\section{\protect\label{S_HIG} The High Temperature Expansion}

We have explained that we will construct the model based on the random
couplings  $\widehat{U}$ by requiring that the high $T$ expansion is the same
than in the original model with complex frustration (and no disorder). Let us
remark that both these models, the random one and the deterministic one, are
{\em regular}, i.e. there are no  couplings of $O(1)$ when $D \to \infty$. In
other words  all the $U$ couplings and the $\widehat{U}$ ones, after being
multiplied times the appropriate $c(D)$ factor, go to zero in this  limit.
Under this condition the high temperature expansion for the XY model (defined
in (\ref{E_XYCON},\ref{E_XYHAM})) is equal to the one of the spherical
model ((\ref{E_SPHCON},\ref{E_SPHHAM})). One can verify this statement by
checking that in the two cases (i.e. for the spherical and for the XY
model) the same diagrams survive in the $D\to\infty$ limit. The regularity
condition guarantees the absence of diverging couplings which could break the
equivalence.

Thanks to this result we will be able to start from computing the high $T$
expansion of  the spherical model ((\ref{E_SPHCON},\ref{E_SPHHAM})), in order
to work out results valid for the XY model (which is the one we study
numerically). That will make our task far easier.

We introduce the Laplacian operator $ \gD$ defined as

\be
  (\gD f)_j \equiv  \sum_k U_{j,k} f_k\ ,
\ee
We denote its spectral density  by $\gr_\gD(\gl)$, and we express the trace
of its $n$-th moment as

\be
  2^{-D}\  \Tr(\gD^n) =\int d\gl\  \gr_\gD(\gl)\  \gl^n\ .
\ee
Here the trace is taken over a space of dimensionality $2^D$, and the
normalizing factor $2^{-D}$ is such that the spectral density of the
identity operator $\gr_1(\gl)$ is $\gd(\gl-1)$.

We start by remarking that the internal energy density of the Gaussian model
is  given in terms of $\gr_\gD(\gl)$ by

\be
  \protect\label{E_EGAUSS}
  E_G= \int\ d\gl\  \gr_\gD(\gl)\  \frac{\gl}{ 1 - \beta \gl}\ .
\ee
By using the expression of the Hamiltonian which includes the spherical
constraint, (\ref{E_SPHHCO}), we see that analogously to (\ref{E_EGAUSS}) we
find

\be
  \protect\label{E_ESPH}
  E_{S} = \int\ d\gl\ \gr_\gD(\gl)\   \frac{\gl}{ \mu(\gb) - \gb \gl}\ ,
\ee
where $\mu$ is a function of $\beta$. It is fixed by the condition

\be
  \protect\label{E_CONLAM}
  \int  d\gl\   \gr_\gD(\gl)\ \frac{1}{ \mu(\gb) - \beta \gl} =1\ ,
\ee
which tells that $\lan \sum_i |\gs_i|^2 \ran=N$, i.e. that the $\gs$
variables satisfy the spherical constraint (\ref{E_SPHCON}).

Equations (\ref{E_ESPH},\ref{E_CONLAM}) can be written in a more compact form
as

\begin{eqnarray}
  \protect\label{E_FINALMU}
  \mu(\gb) &=&   R(\frac{\beta}{\mu(\gb)})\ ,\\
  \protect\label{ENERGIA}
  E(\gb)   &=& \frac{\mu-1}{\gb}\  ,
\end{eqnarray}
where the function $R$ is given by

\be
  \protect\label{E_DEFR}
  R(z)= \int d\gl\ \gr_\gD(\gl)\  \frac{1}{1 -z \gl} \ .
\ee
One uses (\ref{E_FINALMU}) to determine $\mu$, and inserting it in
(\ref{ENERGIA}) one determines the internal energy density of the system.

The critical temperature $\gb_c^{-1}$ is fixed by the condition that
eq.  (\ref{E_FINALMU}) does not admit a solution for $\gb>\gb_c$,
i.e. is such that

\be
  \protect\label{E_CRITICAL}
  z_c \ R(z_c)=  \beta_c\ ,
\ee
where $z_c$ is the inverse of the largest eigenvalue of $\gD$.

In the limit $D\to\infty$, the function $R(z)$ has been computed in
ref.  \cite{PARISI}\footnote{In an appendix of this note we close a
gap of the proof given in \cite{PARISI}.}.  One finds that

\be
 \protect\label{MAGIC}
  G_n^{(q)} = \int \ d\gl \ \gr_\gD(\gl) \gl^n
  = \lan 0|\cX_q^n|0\ran\ ,
\ee
where $\cX_q$ has been defined in (\ref{E_X}).

It can be
shown \cite{PARISI} that the function $R(z)$ has a singularity of the
form

\be
  R(z)= A\ (z_c^2-z^2)^{\frac{1}{2}}\ ,
\ee
where

\be
  z_c=\frac{\sqrt{1-q}}{2}\ .
\ee
The critical behavior does not depend on $q$.

The coefficients of the Taylor
expansion of $R(z)$ around $z=0$ can  be easily evaluated on a computer. The
time cost of the computation increases as the
square of the  order of the highest coefficient one wants to compute. The
asymptotic behavior of the coefficients for large $z$ is controlled by the
singularity closer to $z=0$. If we define

\be
  R(z)=\sum_n R_n z^{2n}\ ,
\ee
we have that

\be
  \protect\label{E_LIMITE}
  \lim_{n \to \infty} R_n\  z_c^{2n}\  n^{\frac32}
\ee
is finite and is given by

\be
 -\frac{A z_c}{2\pi^\frac12} \ .
\ee
We want now to estimate the function $R(z)$ starting from the knowledge of the
first $N$ coefficients of its expansion around $z=0$.

Let us consider the function

\be
  \protect\label{E_ERRE}
  r(z) \equiv \frac{2(1-(1-z^2)^{\frac12})}{z^2}-1 = \sum_n r_n z^{2n}\ .
\ee
Since for large $n$

\be
  r_n \simeq \frac{n^{-\frac32}}{\sqrt{\pi}}
\ee
eq. (\ref{E_LIMITE}) tells us that for large $n$

\be
  R_n \simeq 2 \ A \ z_c^{-2n+1}\  r_n\ .
\ee
Let us say we have computed the coefficients $R_n$ for $n\le N$. We can use
the two higher orders of the series to estimate $A$ and $z_c$, which we will
denote by $\AN$ and $\zN$. They are determined by the relations

\bea
  R_N     = \AN\  r_N\      z_c^{(N)^{(-2N+1)}}\ ,
  \\
  \nonumber
  R_{N-1} = \AN\  r_{N-1} \ z_c^{(N)^{(-2N-1)}}\ ,
\eea
where $r_N$ is the $N$-th coefficient of the expansion of (\ref{E_ERRE})

\be
  r_N = \frac{\gG(N+\frac12)}{\gG(N+2)\gG(\frac12)}\ .
\ee
Now we assume that $\AN$ and $\zN$ are a good estimate for $A$ and $z_c$, and
that for $n>N$ the coefficients $R_n$ of the function $R(z)$ are

\be
  \protect\label{E_RN}
  R_n = \AN r_n z_c^{(N)^{(-2n+1)}}\ .
\ee
We find that our assumption is equivalent to assume that

\be
  \protect\label{E_S_RITA}
  R(z) = \sum_{n=0,N} (R_n-\AN z_c^{(N)^{1-2n}}) z^{2n} + \AN
  z_c^{(N)} r(\frac{z}{z_c})\ .
\ee
The first $N$ coefficients of the Taylor expansion of this function
are  exactly the $R_n$, since the two terms containing $\AN$ cancel.
The higher order terms of the Taylor expansion of (\ref{E_S_RITA})
are given by the terms (\ref{E_RN}).

We have tried in our computation two large values of $N$, i.e.
$N=3\cdot 10^3$ and $N=3\cdot 10^4$. We have computed the expansion of
$R(z)$ around $z=0$ up to order $N$ in the two cases, and we have
found a very similar estimate for $R(z)$.  In the case where $q=0$ the
function $R(z)$ can be computed exactly (\cite{PARISI}), and it is
given by $r(\frac{z}{2})$. In this case one can compute the exact
expression for  $E(\beta)$.

\begin{figure}
\epsfxsize=400pt\epsffile{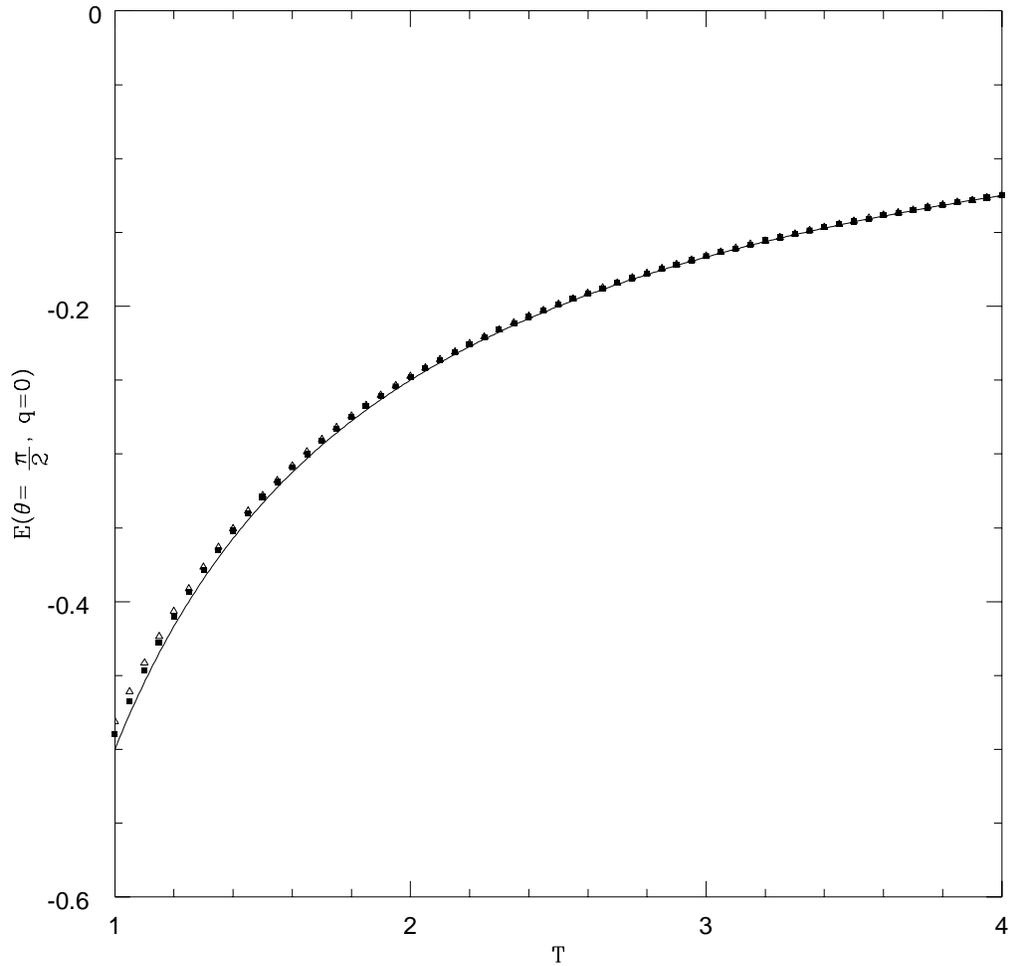}
\protect\caption[1]{Energy of the $q=0$ model versus $T$ in the high
temperature phase. The continuous line is from the resummation of the
high $T$ expansion, the points come from Monte Carlo simulations (for details
see later in the text). In order to give a feeling for the finite size
effects we plot with filled squares the data obtained on our larger
lattice, $D=15$ and $32768$ sites, and with empty triangles data from
a smaller lattice, with $D=12$, i.e. $4096$ sites.
\protect\label{F_EQ0}}
\end{figure}

\begin{figure}
\epsfxsize=400pt\epsffile{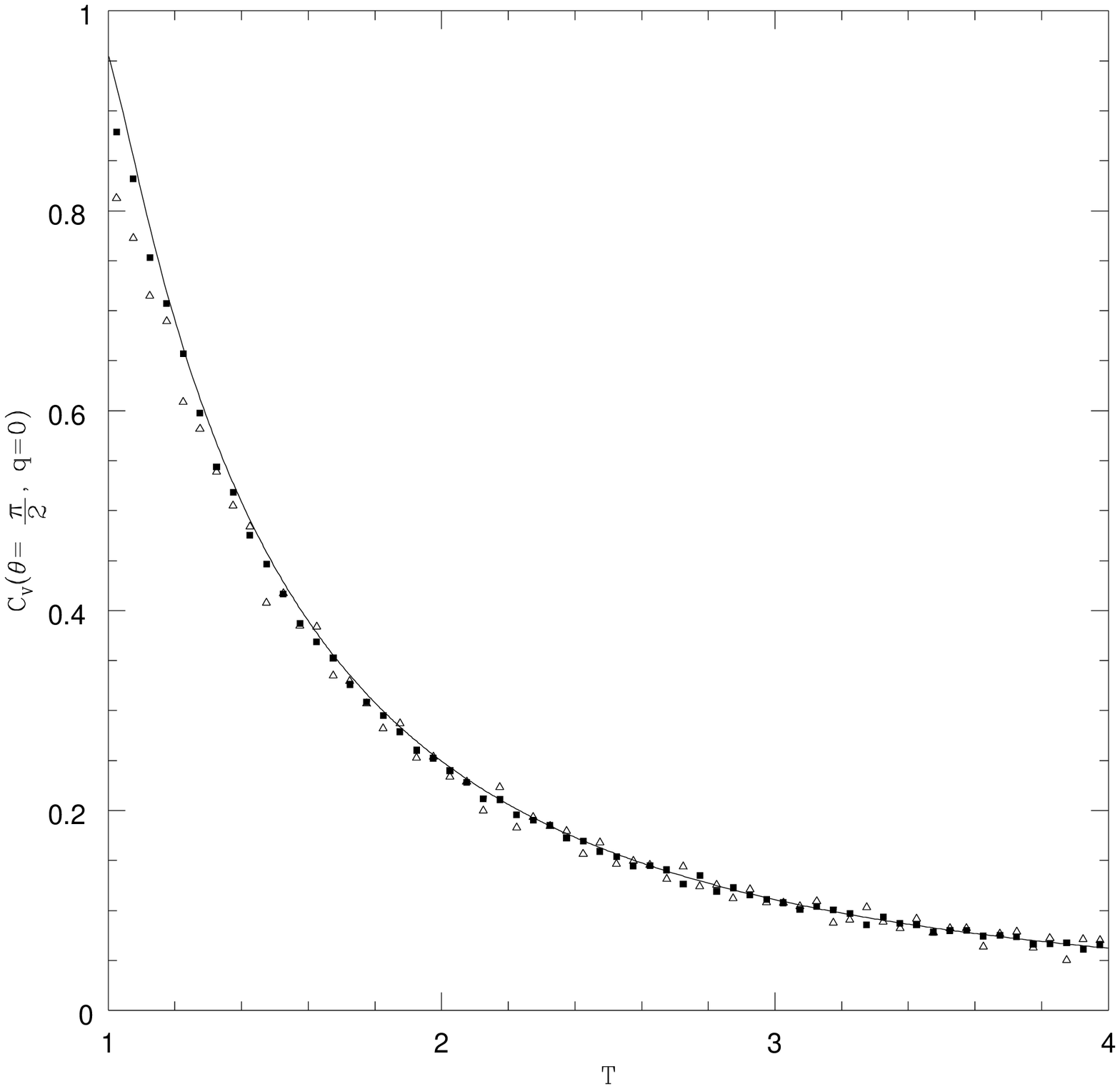}
\protect\caption[1]{
As in fig. (\protect\ref{F_EQ0}) but for the specific heat.
\protect\label{F_CVQ0}}
\end{figure}

We plot $E(\beta)$ and the corresponding specific heat for the case
$q=0$ in figures (\ref{F_EQ0}) and (\ref{F_CVQ0}).  For all values of
$q$ the specific heat at the transition point has the value of $1$.
The figures depict the high $T$ phase, i.e. the region of $T>T_c$
(which we know analytically). The agreement of the Monte Carlo
data (which we will discuss in detail in the following) with the
analytic solution looks quite good, even if on our larger lattice size
we can still distinguish a clear finite size effect.

\begin{figure}
\epsfxsize=400pt\epsffile{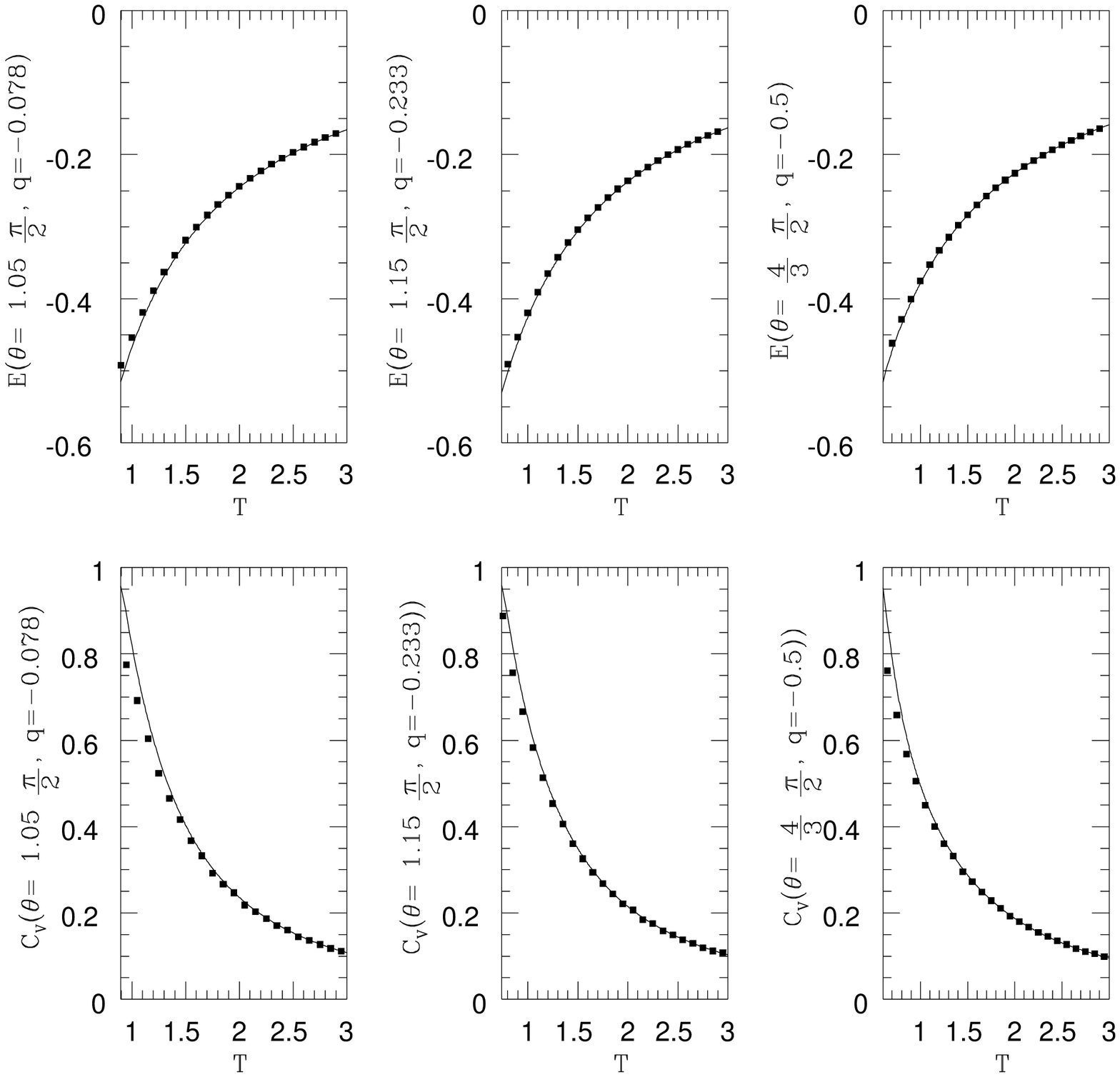}
\protect\caption[1]{Energy and specific heat
of the three models with $q=-0.078$, $q=-0.233$ and $q=-0.5$
versus $T$ in the high
temperature phase. The continuous line comes from the resummation of the
high $T$ expansion, the points are from Monte Carlo simulations (for details
see later in the text). Filled squares are
for the data obtained on our larger
lattice, $D=15$ and $32768$ sites.
\protect\label{F_ECV_QNEG}}
\end{figure}

In fig. (\ref{F_ECV_QNEG}) we plot the energy and the specific heat
for the three cases of  $q=-.078$, $q=-.233$ and $q=-0.5$. The
horizontal scale starts with the critical point. One can observe that
the critical point shifts with $q$. In the specific heat finite size
effects are manifest close to $T_c$.

\begin{figure}
\epsfxsize=400pt\epsffile{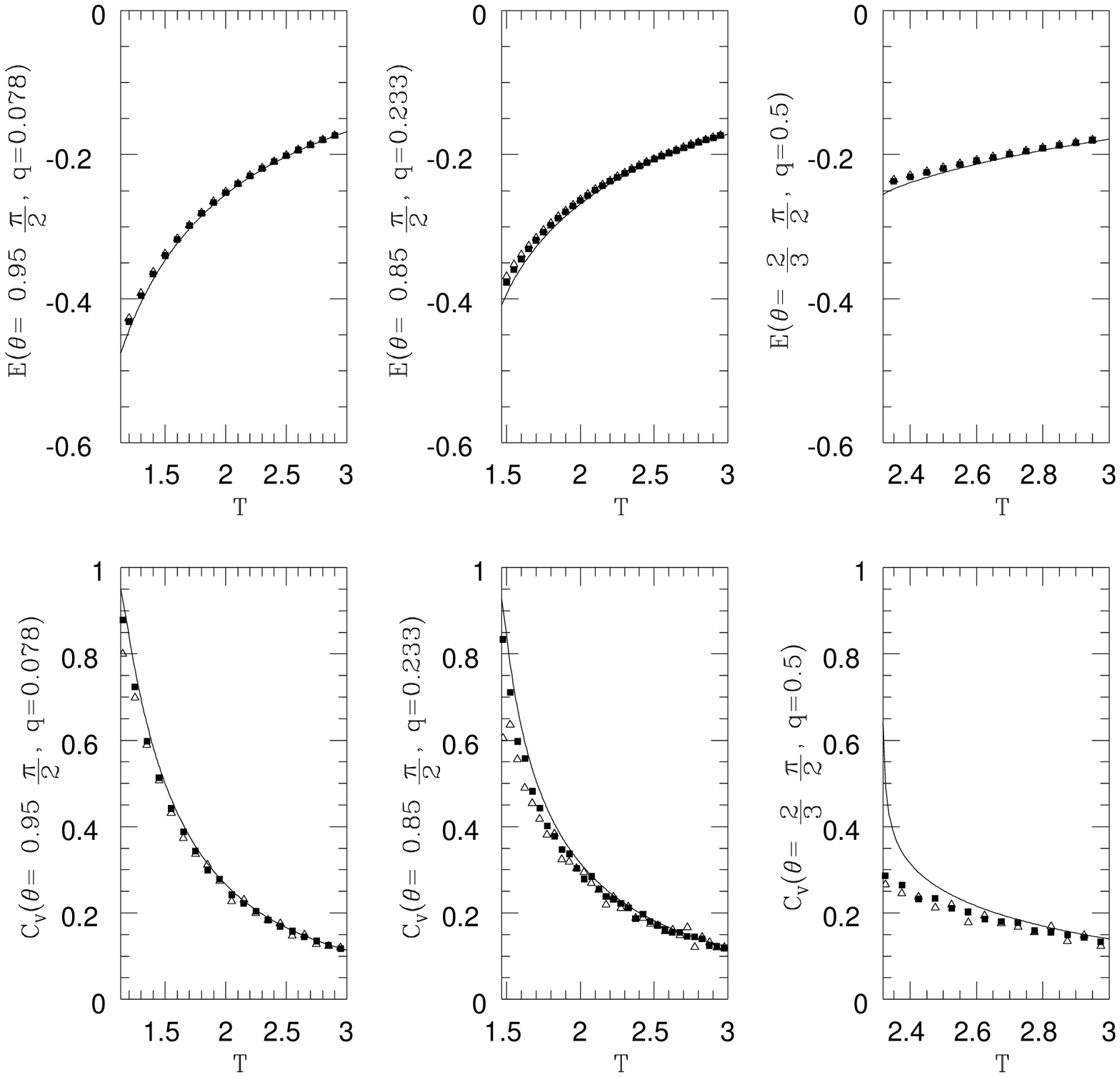}
\protect\caption[1]{As in fig. (\protect\ref{F_ECV_QNEG}), but
for $q=0.078$, $q=0.233$ and $q=0.5$. Here we also add empty triangles
for a smaller lattice size, with the same notation of fig.
\protect\ref{F_EQ0}
\protect\label{F_ECV_QPOS}}
\end{figure}

In fig. (\ref{F_ECV_QPOS}) we plot the energy and the specific heat
for the three analogous cases of positive $q=.078$, $q=.233$ and
$q=0.5$. Here the specific heat has a very sharp variation near the
critical temperature. The variation becomes more and more abrupt for
increasing values of $q$. The situation is dramatic at $q=0.5$. Our
analytic result does not succeed to reproduce the very sharp peak of
the specific heat. We would have needed here a very high accuracy in
order to approximate the correct result. In this case already in the
high $T$ side of the transition the points obtained by numerical
simulations show close to the critical point very ferocious finite
size effects. It is remarkable how non-symmetric around $q=0$ the
situation is. For $q$ negative, i.e. in the direction of the fully
frustrated model, the system is changing quite smoothly. On the
contrary for positive $q$, i.e. when approaching the ferromagnetic
limit, the system changes very drastically. Indeed fig.
(\ref{F_ECV_QPOS}) shows that the change from $q=.233$ to $q=0.5$ is
very dramatic.

We can summarize. A part from the presence of such strong finite size
effects for high positive $q$ the high temperature analysis shows a
very good agreement with the Monte Carlo data, which we will discuss
in better detail in the following.

The spherical approximation is correct in the high temperature phase
also for the model with quenched disorder.  In fact since the
coupling matrices of the disordered model and of the deterministic one
are isospectral the two models coincide in the spherical approximation
and consequently in the high temperature phase.

A last delicate point we want to discuss here is about the
$D\to\infty$ limit.  The reader may wonder about the interchange of
the limit $D \to \infty$ with the limit $\beta \to \beta_c$.  Is that
safe? Couldn't our theorems which allow us to solve the high
temperature phase of the model with complex frustration by using the
$q$-deformed harmonic oscillator be spoiled from such an interchange?
In order to be sure that nothing horrible happens (and also as an
independent check of our numerical simulations) we have computed the
function $R_D(z)$ for a generic value of the dimension $D$ up to the
order $z^{18}$.  This can be done by considering all different (apart
from permutations) closed path of up to $18$ elements, and by
computing their area and multiplicity. Since the total number of
diagrams is $6859315116$ this computation can hardly be done
by hand by simple enumeration. We preferred  to let a computer
to accomplish the task for us.

We define the Taylor series for the finite dimensional function
$R_D(z)$ as

\be
  R_D(z)=\sum_{k=0,\infty} R^k_D z^{2k}\ .
\ee
We also define

\begin{equation}
  q_n \equiv \cos(n \theta)\ ,
\end{equation}
where, obviously, $q_1 = q$.
We give here the full expression for the first $5$ coefficients we
have computed:

\begin{eqnarray}
  \nonumber R^0_D &=& 1 \ ,\\
  \nonumber R^1_D &=& 1 \ ,\\
  \nonumber R^2_D &=& \frac{(D-1)}{D}(2 + q_1) + \frac{1}{D} \ ,\\
  \nonumber R^3_D &=& \frac{(D-2)(D-1)}{D^2}(5 + 6 q_1 + 3 q_1^2 + q_1^3)\\
                &+&  \frac{(D-1)}{D^2}(9 + 6 q_1) + \frac{1}{D^2}\ ,\\
  \nonumber R^4_D &=& \frac{(D-3)(D-2)(D-1)}{D^3}
        (14 + 28 q_1 + 28 q_1^2 + 20 q_1^3 + 10 q_1^4+ 4 q_1^5+ q_1^6) \\
  \nonumber     &+& \frac{(D-2)(D-1)}{D^3}
                (56 + 86 q_1 + 52 q_1^2 + 16 q_1^3) \\
  \nonumber     &+& \frac{(D-1)}{D^3}
                (34 + 28 q_1 + q_2)  + \frac{1}{D^3} \ .
\end{eqnarray}
Let us also define the leading contribution to $R^k_D$ as the terms of order
one which multiply the different powers of $q_1$,
and the first one over $D$ corrections analogously,
i.e.

\be
   R^k_D \equiv
    \sum_{\ga=0} \cR^{k,\ga}q_1^\ga
  \bigl[1-\frac{k(k-1)}{2D}\bigr]
  + \sum_{\ga=0} \frac{\cS^{k,\ga}}{D}     q_1^\ga
   + O(\frac{1}{D^2})\ ,
\ee
since the leading and the subleading terms
in $D$ contains only powers of $q_1$ and not of the
others $q_n$.
In tables (\ref{T_UNO}-\ref{T_NOVE}) we give all the $\cR^{k,\ga}$ and the
$\cS^{k,\ga}$ we have computed. We hope that this information maybe
useful for a possible analytic computation of the $\frac1{D}$ corrections.

\begin{table}
\begin{tabular}{||  c || c|c|c|c|c|c|c ||}\hline\hline
$\alpha$  & 0 & 1 &  2 & 3 & 4 & 5 & 6  \\
k  &  &  &   &  &  &  &   \\ \hline\hline
0 & 1  &   &    &   &   &   &    \\
1 & 1  &   &    &   &   &   &    \\
2 & 2  & 1 &    &   &   &   &    \\
3 & 5  & 6 & 3  & 1 &   &   &    \\
4 &14  &28  &28   &20  & 10  &4   &1    \\
5 &42 & 120 &  180&    195 &  165 &      117 &   70 \\
6 &   132  &   495 &  990  & 1430  &  1650  &  1617  &   1386  \\
7 & 429  &  2002  &   5005  & 9009  & 13013   &  16016  &  17381   \\
8 & 1430 &   8008 &  24024  &  51688 &  89180  &  131040   &  169988   \\
9  &  4862  & 31824  & 111384 &  278460  &    556920 &  946764 & 1419432\\
\hline\hline
\end{tabular}
\protect\caption[a]{The coefficients ${\cal R}^{k,\alpha}$ for $\alpha$ going
from $0$ to $6$.
\protect\label{T_UNO}}
\end{table}

\begin{table}
\begin{tabular}{||  c || c|c|c|c|c|c ||}\hline\hline
$\alpha$  & 7 & 8 &  9 & 10 & 11 & 12  \\
k  &  &  &   &  &  &   \\ \hline\hline
5  &     35     &  15    &    5       &   1         &         &        \\
6  &  1056      & 726    &   451  &      252  &   126    &       56    \\
7  & 16991    &  15197   &  12558  &   9646   &   6916   &     4641    \\
8  &   199264 &  214578  &  214760 &  201460    &   178248 & 149464 \\
9  &  1922904 & 2394450 & 2775080 & 3021444 & 3112632& 3051024\\
  \hline\hline
\end{tabular}
\protect\caption[a]{As in table
(\protect\ref{T_UNO}) but for $\alpha$ going from
$7$ to $12$.
\protect\label{T_DUE}}
\end{table}

\begin{table}
\begin{tabular}{||  c || c|c|c|c|c|c ||}\hline\hline
$\alpha$  & 13 & 14 &  15 & 16 & 17 & 18  \\
k  &  &  &   &  &  &   \\ \hline\hline
6  &   21   &     6   &   1   &    &    &     \\
7  &   2912   &    1703    &  924    &  462   & 210     & 84      \\
8  &  119168 &  90540 & 65640 & 45438 & 30024   &  18908   \\
9  & 2858040  &  2567340 & 2217480  &  1845486  & 1482264 & 1150220\\
\hline\hline
\end{tabular}
\protect\caption[a]{As in table
(\protect\ref{T_UNO}) but for $\alpha$ going from
$13$ to $18$.
\protect\label{T_TRE}}
\end{table}

\begin{table}
\begin{tabular}{||  c || c|c|c|c|c|c|c ||}\hline\hline
$\alpha$  & 19 & 20 &  21 & 22 & 23 & 24 & 25  \\
k  &  &  &  &  &  & &   \\ \hline\hline
7  &  28   &  7  &  1   &    &    &    &    \\
8  &  11320  & 6420   &  3432   & 1716   & 792 &330 & 120 \\
9  &  862920 &    626076 & 439263  & 297891    &  195075  &   123165 & 74817\\
\hline\hline
\end{tabular}
\protect\caption[a]{As in table
(\protect\ref{T_UNO}) but for $\alpha$ going from
$19$ to $25$.
\protect\label{T_QUATTRO}}
\end{table}

\begin{table}
\begin{tabular}{||  c || c|c|c|c|c|c|c|c|c|c|c  ||}\hline\hline
$\alpha$  & 26 &  27 & 28 & 29 &   30 &   31 &  32 &  33 & 34 & 35 & 36\\
k  & &   &  & &  &  &  & & & & \\ \hline\hline
8  &  36  & 8 & 1 &  &  &  &
      &  &  &   &  \\
9  &  43605  & 24293 & 12870 & 6435 & 3003 & 1287 &
     495 & 165 & 45 &  9 &  1\\
\hline\hline
\end{tabular}
\protect\caption[a]{As in table
(\protect\ref{T_UNO}) but for $\alpha$ going from
$26$ to $36$.
\protect\label{T_CINQUE}}
\end{table}

\begin{table}
\begin{tabular}{||  c || c|c|c|c|c|c|c ||}\hline\hline
$\alpha$  & 0 & 1 &  2 & 3 & 4 & 5 & 6  \\
k  &  &  &   &  &  &  &   \\ \hline\hline
2 & 1  &  &    &   &   &   &    \\
3 & 9  & 6 &   &  &   &   &    \\
4 &56  &86  &52   &16  &   &   &    \\
5 &300 & 740 &  880&    690 &  370 &      140 &   30 \\
6 & 1485  &5082   & 8904  &10818  &10020   &7494  & 4611  \\
7 &  7007  & 30758 & 70707  & 114471  &145264  & 153377  & 139286  \\
8 &  32032 &171808 & 486920 & 976520  &1548952 & 2064048 &2395464    \\
9 &  143208 &908208   &3052656 &7265664 &13712319 & 21806163 &30323493 \\
\hline\hline
\end{tabular}
\protect\caption[a]{The coefficients ${\cal S}^{k,\alpha}$ for $\alpha$ going
from $2$ to $6$.
\protect\label{T_SEI}}
\end{table}

\begin{table}
\begin{tabular}{||  c || c|c|c|c|c|c ||}\hline\hline
$\alpha$  & 7 & 8 &  9 & 10 & 11 & 12  \\
k  &  &  &   &  &  &   \\ \hline\hline
6  & 2310 &927 &276 &48&       &        \\
7  & 110691& 77882   & 48727  & 26964  &13020 & 5397    \\
8  & 2476448 & 2316576 & 1981972 & 1560904    & 1135608 & 764856 \\
9  & 37776564 &42883740 &44909478 & 43774344 & 39972618 & 34364322\\
  \hline\hline
\end{tabular}
\protect\caption[a]{As in table
(\protect\ref{T_SEI}) but for $\alpha$ going from
$7$ to $12$.
\protect\label{T_SETTE}}
\end{table}

\begin{table}
\begin{tabular}{||  c || c|c|c|c|c|c ||}\hline\hline
$\alpha$  & 13 & 14 &  15 & 16 & 17 & 18  \\
k  &  &  &   &  &  &   \\ \hline\hline
7  &   1848   &    476    &  70    &     &      &       \\
8  &  476704 &  273784 &143804 & 68424 & 29116   &  10800   \\
9  &  27912096  &  21466764 & 15650046  &  10819422  & 7090146 & 4396734\\
\hline\hline
\end{tabular}
\protect\caption[a]{As in table
(\protect\ref{T_SEI}) but for $\alpha$ going from
$13$ to $18$.
\protect\label{T_OTTO}}
\end{table}

\begin{table}
\begin{tabular}{||  c || c|c|c|c|c|c|c|c|c|c ||}\hline\hline
$\alpha$  & 19 & 20 &  21 & 22 & 23 & 24 & 25 & 26 & 27 & 28  \\
k  &  &  &  &  &  & & & & &  \\ \hline\hline
8  & 3312 & 752   &  96  &    &  & & & & &  \\
9  & 2571534 & 1411929 & 723609& 343611& 149490& 58410 &19800&5490&1116&126 \\
\hline\hline
\end{tabular}
\protect\caption[a]{As in table
(\protect\ref{T_SEI}) but for $\alpha$ going from
$19$ to $28$.
\protect\label{T_NOVE}}
\end{table}

For analyzing the large $D$ behavior of our series is useful to define
the expansion

\be
  R(z) = \sum_k z^{2k} \sum_\gd \overline{R}^k_\gd D^{-\gd}\ ,
\ee
where the $\gd=0$ contribution is the leading term of the $D^{-1}$
expansion. We define the quantity

\be
   \gO^k(\gth) \equiv \frac{\overline{R}^k_0}{\overline{R}^k_1}\ ,
\ee
which is related to the convergence radius of the $k$-th term in
$D^{-1}$. It is indeed easy to see that in the large $D$ and large $k$
limit

\be
  \gO^k(\gth) \simeq k(\cC(D)-\cC(\infty))\ ,
\ee
where $\cC(\infty)$ is the radius of convergence of the perturbative
series in
$D=\infty$, and $\cC(D)$ is the radius of convergence of the series in a
finite number of dimensions $D$.

\begin{figure}
\epsfxsize=400pt\epsffile[28 550 566 811]{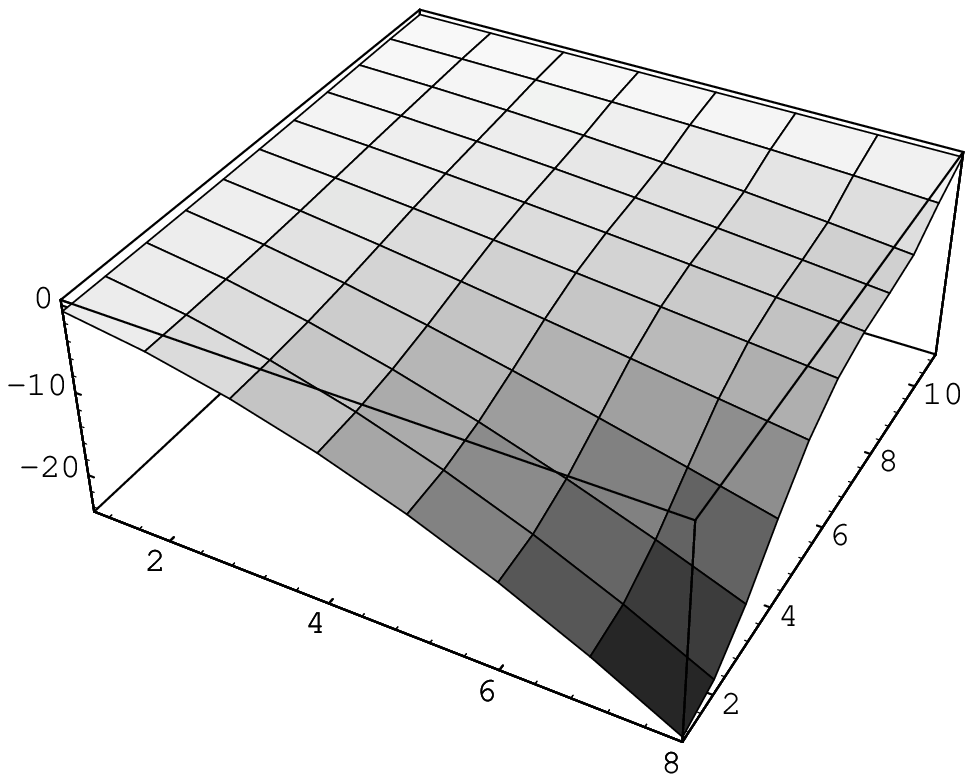}
\protect\caption[1]{
$\gO^k$ versus for $k$ going from $2$ to $9$. The axis labeled with
$2$, $4$, $6$ and $8$, on the left,  is $k-1$.
The axis with labels going up to $10$, on the right,
is the $\gth$ axis. $\gth$ goes from $0$ to $\pi$. $\gth=0$ coincides
with the tick $1$, $\gth=\frac{\pi}{2}$ with the tick $6$ and
$\gth=\pi$ with the tick $11$.
$q=1$ on the left limit of the axis (ferromagnet), $q=0$ (spin glass)
in the center
and $q=-1$ (fully frustrated model) on the right edge of the axis.
The vertical axis is for $\gO^k$.
Decreasing values of $\gO^k$ are drawn with darker coloring.
\protect\label{F_3D1}}
\end{figure}

\begin{figure}
\epsfxsize=400pt\epsffile[28 550 566 811]{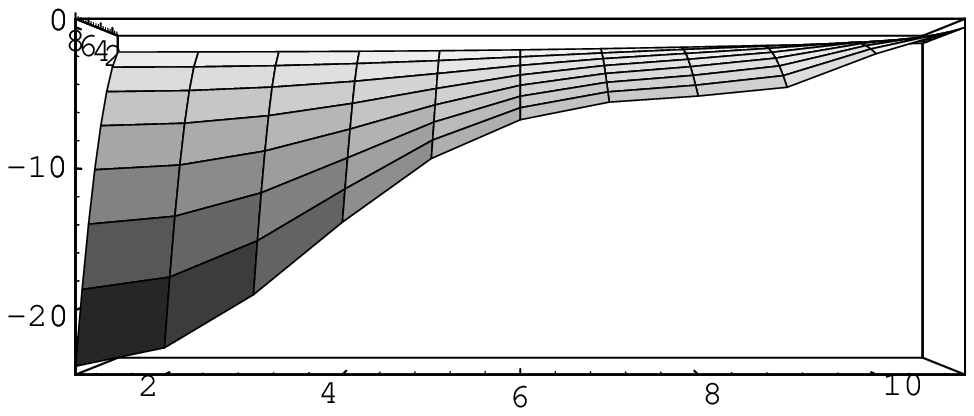}
\protect\caption[1]{
As in fig. (\protect\ref{F_3D1}), but from a different point of
view. Here the $x$ axis is $\gth$ ($0$ on the left and $\pi$ on the
right), the $y$ axis is $\gO$ and the $k$ axis goes beyond the page.
We sit on the side of high $k$ values.
\protect\label{F_3D2}}
\end{figure}

\begin{figure}
\epsfxsize=400pt\epsffile[28 550 566 811]{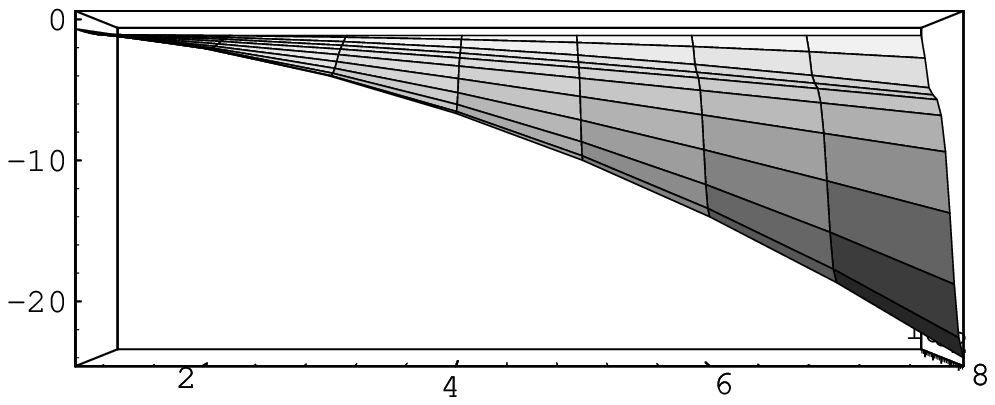}
\protect\caption[1]{
As in fig. (\protect\ref{F_3D1}), but from a different point of
view. Here the $x$ axis is $k$ ($2$ on the left and $9$ on the
right), the $y$ axis is $\gO$ and the $\gth$ axis goes beyond the page.
We sit on the side of $\gth\simeq 0$ values.
\protect\label{F_3D3}}
\end{figure}

We plot the $\gO$ surface as a function of $k$ and $\gth$ from
different viewpoints in figures (\ref{F_3D1}-\ref{F_3D3}). It is
interesting to note that moving away from $\gth=\frac{\pi}{2}$ in the
direction of the fully frustrated model, i.e. increasing $\gth$, $\gO$
changes quite smoothly. On the contrary when $\gth$ becomes smaller
than  $\gth=\frac{\pi}{2}$ the change is far more abrupt. This is
coherent with what we find from fig. (\ref{F_ECV_QNEG}) where for
negative values of $q$ the system does not change much, and fig.
(\ref{F_ECV_QPOS}) where $q>0$ the system undergoes a strong
quantitative change around $q=\frac12$. This point is where in
figures (\ref{F_3D1}-\ref{F_3D3}) we can find a maximal change of
$\gO$ as a function of $q$. The plateau that one sees when moving to the
right of fig. (\ref{F_3D2}) can be seen in perspective in  fig.
(\ref{F_3D3}).

We have also checked that the results are in reasonable agreement with
the ansatz

\be
  \frac{R^k_D}{R^{k-1}_D} =a +\frac{b(\frac{k}{D})}{D}\ ,
\ee
where the function $b(w)$ does not seem to have a fast divergence for large
values of $w$. Apparently the limit $D \to \infty$ is smooth.

\section{\protect\label{S_LOW}The Low Temperature Region}

In the former section we have discussed the high $T$ region of the
deterministic model with complex frustration.  We have shown that the Monte
Carlo data reproduce well (but for the case of high, positive $q$, where
finite size effects are dramatic) the series obtained by computing the
Green functions of the $q$-deformed harmonic oscillator.  Together with the
results of (\cite{PARISI}) and the Appendix of this note that makes the
status of the high $T$ phase clear.  We also know that in the high $T$
phase the model with quenched disorder coincides by construction with the
deterministic model, but we will see that better in the following.

In order to get information about the low $T$ phase we have to use the
random model, which we have defined in eqs. (\ref{E_RANA},\ref{E_RANB}).
We will use replica theory to solve it both in the high $T$ phase (where
we will find again the same high $T$ series) and in the low $T$ phase.
We will try to understand how much the replica formulation of the system
is connected to the Monte Carlo data we will get directly from the
deterministic model with complex frustration.

Let us solve the random model by using the techniques  introduced in
\cite{MAPARI_1,MAPARI_2}. The computation follows quite closely the one of
\cite{MAPARI_1,MAPARI_2}, and we will give here only the main details.
One introduces $n$ replicas, where $n$ has to be sent to zero at the end of
the computation. The $n$-dependent free energy is given by

\be
  f^{(n)}(\beta) \equiv  - \lim_{N \to \infty} {1 \over \beta N}
  \frac{\overline{Z_U^n}-1}{n}\ ,
\ee
where the bar denotes the average over the random couplings and the
replicated
partition function $Z_U^n$ depends over the noise and can be written as

\be
  Z_U^n \equiv \int [d\gs] e^{ -\beta \sum_{a=1}^n H_U^a}\ .
\ee
The integration over the unitary group can  be done explicitly.
After some algebra one finds that one has to evaluate the stationary
points of the following free energy:

\be
  A[Q,\gL]= -\Tr G(\beta Q)+ \Tr (\gL Q) -F(\gL)\ ,
  \protect\label{eqA}
\ee
where $Q$ and $\gL$ are $n \times n$ matrices, the function $G$ is
related to the one defined in eq. (\ref{ENERGIA}) by

\be
  \protect\label{E_DGDZ}
  {dG \over dz} \equiv E(z)\ ,
\ee
and

 \be
F(\gL) \equiv \ln \int d [\gs] \exp( \sum _{a,b}\gL_{a,b} \gs^a \gs^b)\ .
\ee
In the high temperature phase the off-diagonal terms of the two matrices
$Q$ and $\gL$ are zero. If we set

\bea
  Q_{a,b}   &=& \gd_{a,b}\  q  \ ,    \\
  \gL_{a,b} &=& \gd_{a,b}\  \gl\ ,
\eea
we find that the stationary equations imply that

\be
  q = 1 \ \mbox{and}\  \gl=E(\beta)\ .
\ee
We finally find that in the high temperature phase

\be
  {\partial F \over \partial \beta} = E(\beta)\ ,
\ee
where $E(\gb)$ is the function defined in (\ref{E_DGDZ}).  In this way
we have derived again the equivalence of the model with quenched
disorder and the deterministic model with complex frustration in the
high temperature phase.

In the low temperature region the off-diagonal terms of the two matrices are
non-zero. If we assume that replica symmetry is unbroken, we have
that the off-diagonal terms\footnote{We set $Q_{a,a}$=1. The value we
choose for $\gL_{a,a}$ is irrelevant, and does not change the results.}
are given by

\bea
  Q_{a,b}   &=&  q \ ,\\
  \gL_{a,b} &=&  \gl\ .
\eea
In this way we find that we have to minimize the free energy

\be
  \protect\label{E_LOWT}
  G(\beta(1-q))+\beta q E( \beta(1-q)) -\gl q + f(\gl)\ ,
\ee
where the function $f$ is given by

\be
  \ln \bigl( \int dh \exp (-h^2/2) \bigr)
  \ln \bigl(\int d\gs_r d\gs_i \gd(\gs_r^2+\gs_i^2-1)
  \exp (-\gl^{\frac{1}{2}} h \gs_r)\bigr)\ .
\ee
The energy turns out to be

\be
E(\gb)=G'(\gb(1-q))-\gb q(1-q) G''(\gb(1-q))\ .
\ee
By deriving this expression and evaluating it for $\gb=\gb_c$ we find that

\be
  C_V(\gb_c^+) = C_V(\gb_c^-) = 1\ .
\ee
The critical temperature can also be determined through the relation

\be
  \gb_c^2 G''(\gb_c) = 1\ .
\ee
One also finds that at zero temperature

\be
  C_V(\infty) = \frac12\ ,
\ee
in agreement with the equipartition theorem.

The equations which determine the minimum of such free energy can be
solved numerically.

We will show and discuss their solution in next section, for different
$q$ values, together with the Monte Carlo results in the low $T$
phase.

We expect the unbroken replica solution to give rather accurate values
for the free energy.  In the SK model the error over the correct,
replica broken result is smaller than $3 \%$, and it is likely to be
even smaller in the present case.  It is interesting to note that the
replica symmetric solution normally gives a lower bound to the true
free energy and to the true internal energy of the system.  Our
numerical simulations show that when we compare numerical simulations
of the deterministic model to the replica symmetric solution of the
disordered model in the cold phase this is not always the case in our
system, pointing to a non complete coincidence of the two models.

\section{\protect\label{S_COM}Computer Simulations}

We will describe here our numerical simulations of the model with
complex frustration and no quenched disorder (but for the small one
needed for constructing the antisymmetric tensor $\cS$), defined with
the couplings of eq.  (\ref{E_STA}), and compare them with the
analytic solution of the model with quenched disorder that we have
discussed in the previous section.  Here we will mainly focus on the
low $T$ phase.

We have simulated systems with $D$ going from $3$ to $15$ or $16$,
i.e.  containing from $8$ to $32768$ or $65536$ sites.  We have been
starting from all fields set to $1$ at high $T$, and decreased the
temperature in small steps.  A typical pattern has been starting from
$T=4.05$, and decreasing it down with $80$ steps of $\gD T = .05$ (but
for some runs we only used $40$ steps and a lower starting point).  At
each next $T$ we have been continuing from the last configuration
obtained at $T+\gD T$.  At each $T$ value we have used $500$ full
sweeps of the system to obtain an acceptance value of the Monte Carlo
procedure of $50\%$ (by tuning the angular increment we would propose
for updating the field phase in a given site).  After that we have
used $1250$ full sweeps to thermalize the system, and $5000$ full
sweeps to measure the internal energy.  We have ran some longer
simulations to check we have indeed reached thermal equilibrium, and
it seems to be the case.  We believe that the statistical error on our
data points is always smaller than the symbols we use to plot them.
We have always only used in the final plots the data from a single
realization of the antisymmetric tensor $\cS$ (even if we have checked
the size of typical fluctuations by simulating more than one $\cS$
set, and the induced uncertainty turned out to be not very large, but
detectable).

As a first check we have verified we could reproduce the results obtained
in \cite{MAPARI_FF} for the fully frustrated model.

A second preliminary question was concerning the equality of the traces
of the $n$-th powers of the coupling matrix and the expectation values of
the operators which appear in the formalism of the $q$-deformed harmonic
oscillator. This is a point which has been proven in (\cite{PARISI}) and
in this note, and verifying it was meant to constitute both a check of
our codes and of our theorems. So given the couplings we have selected,
according to eq.  (\ref{E_STA}) and to a random choice of the $\cS$
(over which in this case we have averaged) we have verified that

\be
  2^{-D} \Tr(\gD_q^n) = \bra \cX^n_q \ket \ .
\ee
In this case we have kept the statistical error (given in this case by the
distribution of the $\cS$, and not by a Monte Carlo: there is no Monte
Carlo here!) under careful control. All momenta up to $n=8$ coincide with
the $q$-deformed result with a better precision than $10^{-3}$. Our best
fits give the right answer, with a $\chi^2$ of order one per degree of
freedom.  So, this check has been positive, and it is an important check
of the equivalence of the model with complex frustration and the random
model in the high $T$ phase.

We come now to the main point of our investigation, i.e. the low $T$
phase. Here we will compare the analytic solution\footnote{We will use the
replica symmetric solution, which we
believe is not too wrong, as we have explained in the former section.} of
the random model (\ref{E_RANA}) with the numerical simulation of the
deterministic model. We will see that the data are indicative of a strong
similarity, but not of a complete equivalence of the two models.

\begin{figure}
\epsfxsize=400pt\epsffile{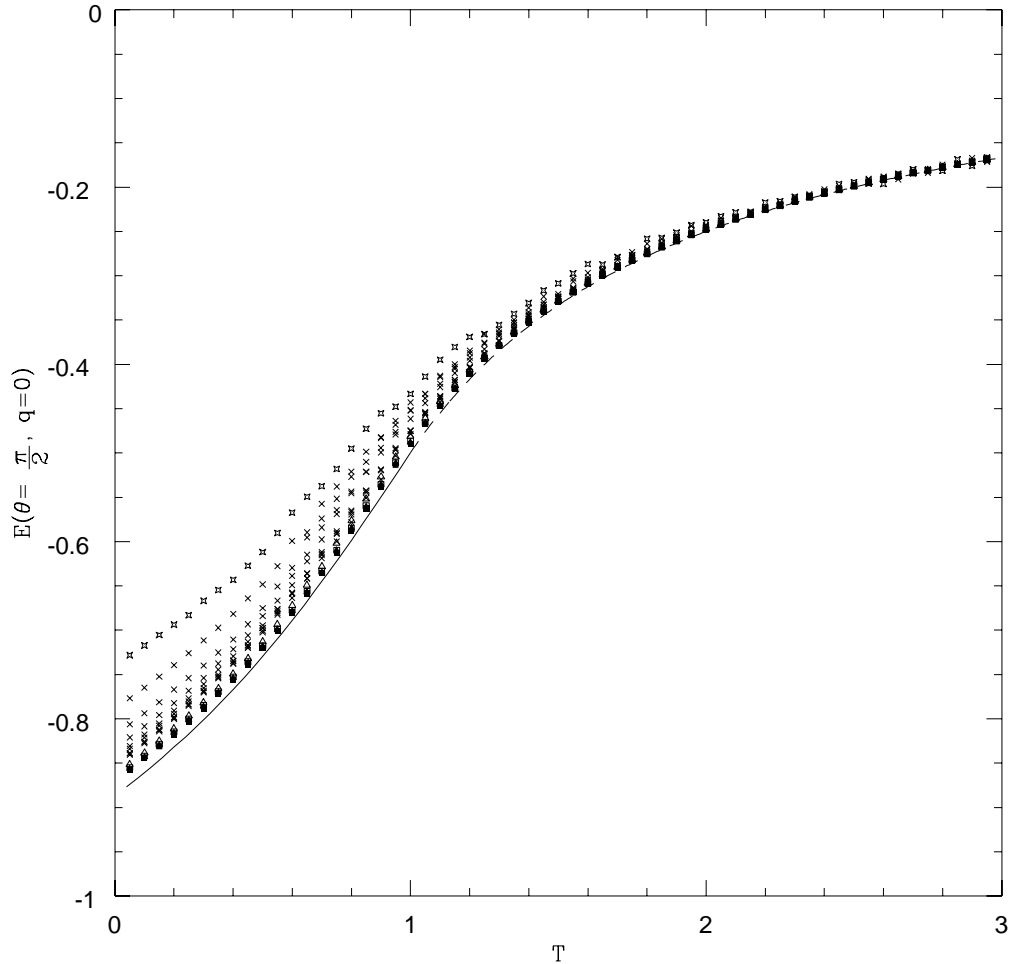}
\protect\caption[1]{
Energy of the $q=0$ model versus $T$. Here we are
looking at both phases (the critical point is at $T_c=1$).
See the text for the explanation of the different symbols.
The point where the dashed line becomes continuous is here and in the
following figures the critical point.
\protect\label{F_EQ0ALL}}
\end{figure}

In fig. (\ref{F_EQ0ALL}) we plot the energy of the $q=0$ model versus
$T$, in both phases (the critical point is at $T_c=1$).  The point
where the dashed line becomes continuous is here and in the following
figures the critical point.  We plot the analytic result from the high
$T$ expansion with a dashed line, while the result obtained by
minimizing eq. (\ref{E_LOWT}) is plotted with a continuous line, for
$T<T_c$. Here we include the data from all our simulations. The fancy
starred dots, lying at the top, are from $D=3$. Crosses are for
intermediate values of $D$ (lower points for higher $D$ values). For
the four higher values of $D$ (in this case $D=13$, $14$, $15$ and
$16$) we change symbol again, and we use respectively empty triangles,
empty squares, filled triangles and filled squares.

The agreement of Monte Carlo data for the deterministic model and
replica symmetric solution of the random model is quite good also in
the broken phase, for $T<T_c$. We expect that the solution with
broken replica symmetry will have an energy slightly higher than the
unbroken one (as we already said, in the general case the replica
symmetric energy is a lower bound to the true energy of the physical
system). Very small residual finite size effect, and this small energy
drift to the the breaking of replica symmetry should explain the small
discrepancy between the numerical data and the analytic curve. So in
the case of the $q=0$ model things seem to go smoothly.

When moving on the side of negative $q$ values things do not change
much, and if there is a discrepancy it is very small. This is
completely coherent with the discussion of the behavior of the
coefficients of the high $T$ expansion of former section.

\begin{figure}
\epsfxsize=400pt\epsffile{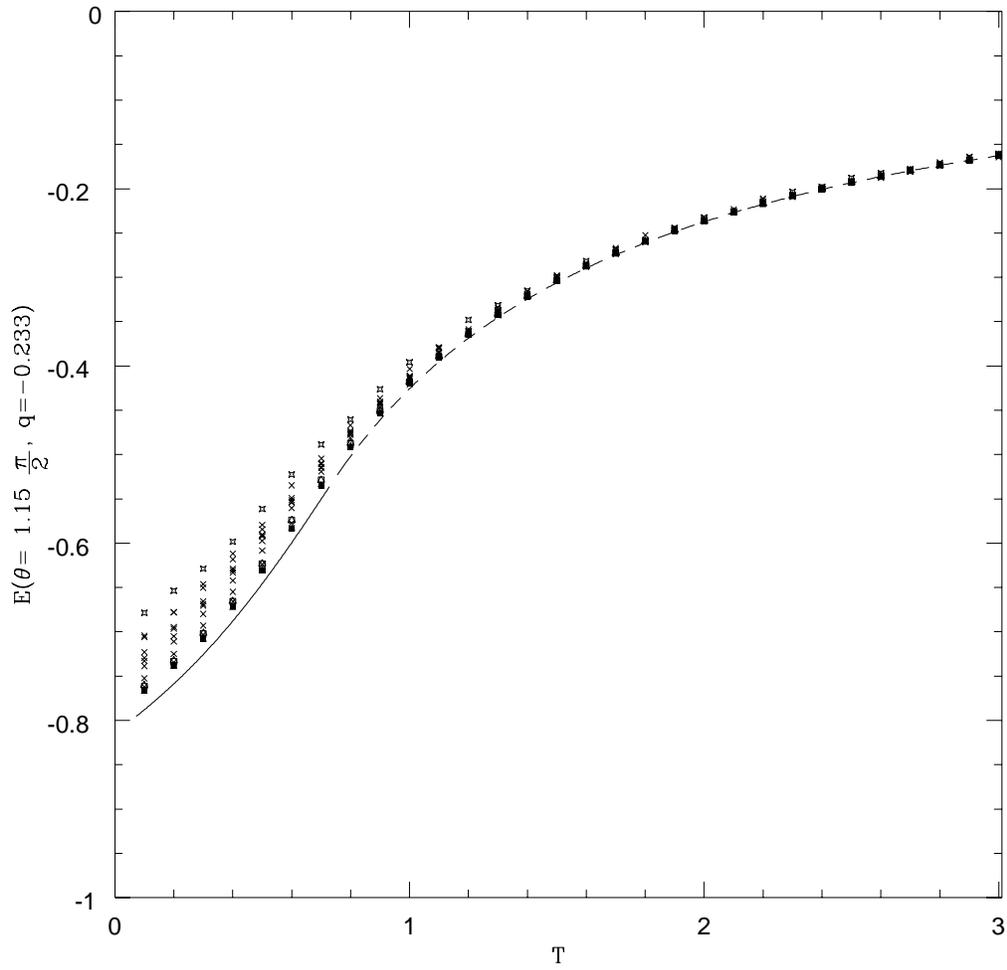}
\protect\caption[1]{
As in fig.  (\protect\ref{F_EQ0ALL}), but for $q=-.233$.
\protect\label{F_EQM233ALL}}
\end{figure}

\begin{figure}
\epsfxsize=400pt\epsffile{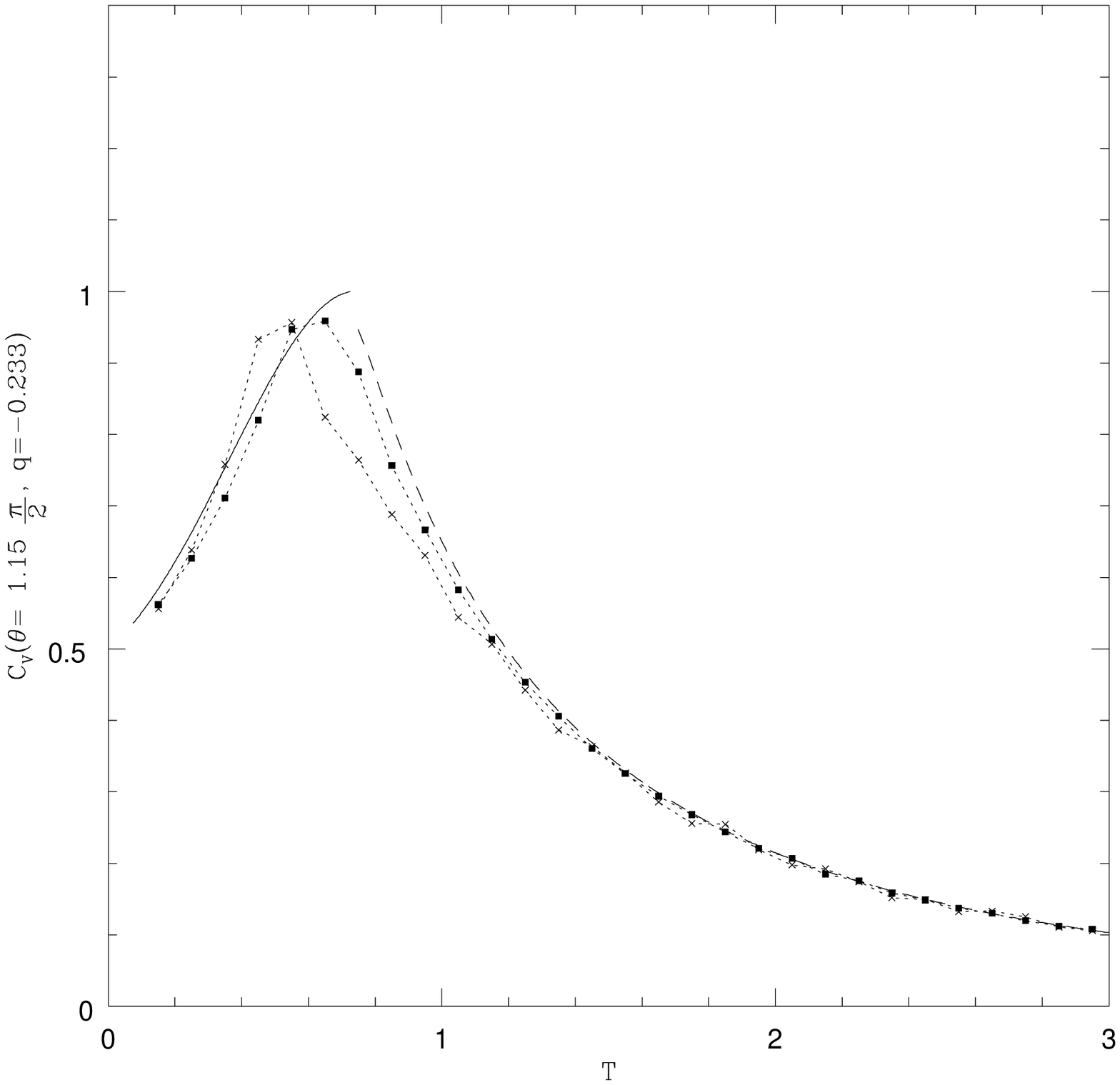}
\protect\caption[1]{
As in fig.  (\protect\ref{F_EQ0ALL}), but the specific heat $C_V$
for $q=-.233$. For sake of graphical clarity here we only use crosses
for $D=12$ data and filled squares for $D=16$, and we join the data
points with dotted lines.
\protect\label{F_CVQM233ALL}}
\end{figure}

In fig. (\ref{F_EQM233ALL}) the results for $q=-.233$ (where the angle
is already different of a $15 \%$ from $\gth=\frac{\pi}{2}$). The
agreement of our data with the analytic solution are still quite
good. Fig. (\ref{F_EQ0ALL}) and fig. (\ref{F_EQM233ALL}) are on the
same scale (as it will be for all the following energy plots). That
allows the reader to appreciate that indeed the two energy plots are quite
different among them. To show even better that things are basically
working in this regime of negative $q$ values we plot
in fig.  (\ref{F_CVQM233ALL}) the specific heat
for $q=-.233$. The small gap in the analytic curve close to the
maximum is because we stopped early our numerical evaluation of the high
$T$ series. We cannot detect here any clear discrepancy.

\begin{figure}
\epsfxsize=400pt\epsffile{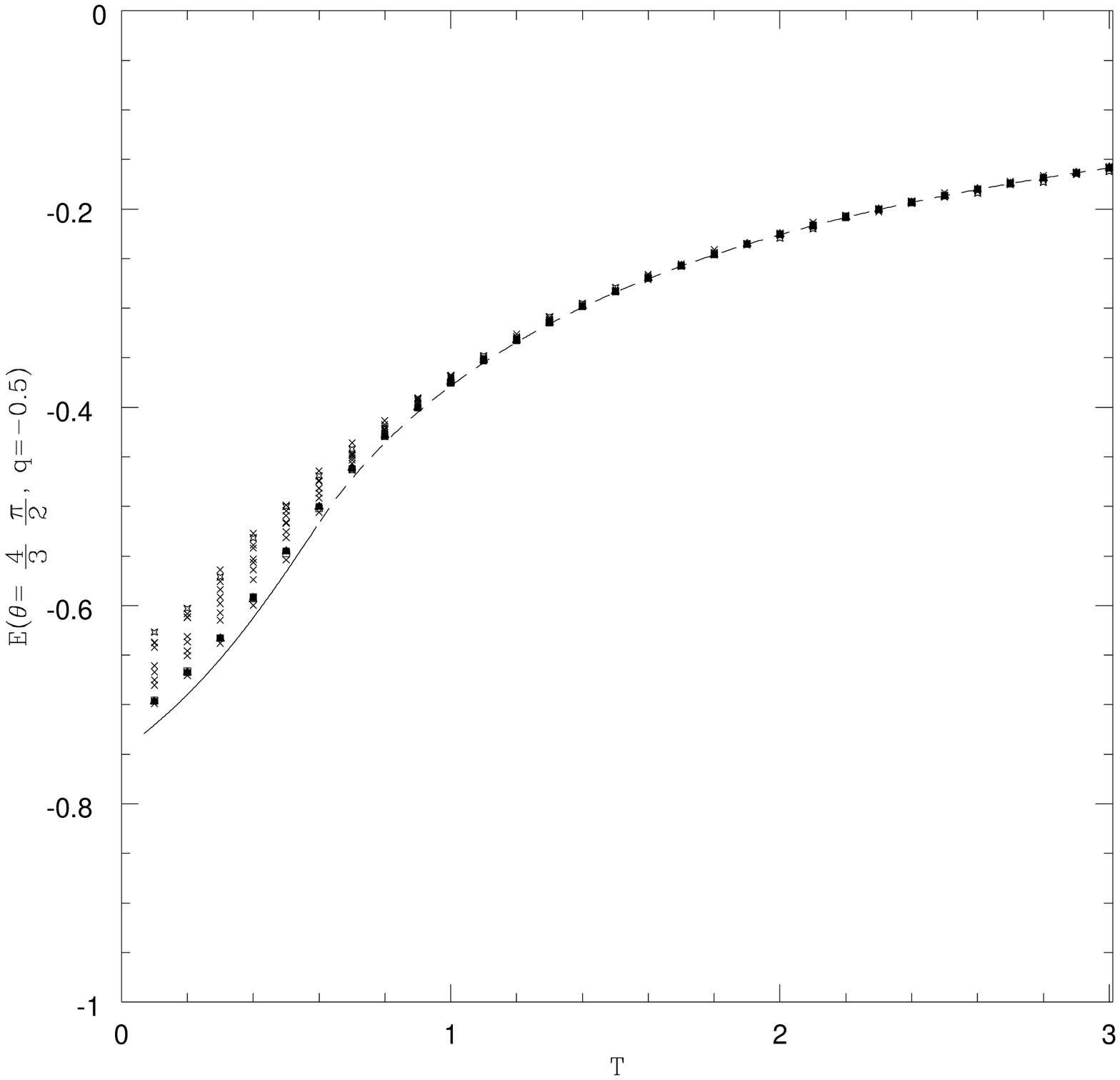}
\protect\caption[1]{
As in fig.  (\protect\ref{F_EQ0ALL}), but for $q=-.5$.
\protect\label{F_EQM5ALL}}
\end{figure}

\begin{figure}
\epsfxsize=400pt\epsffile{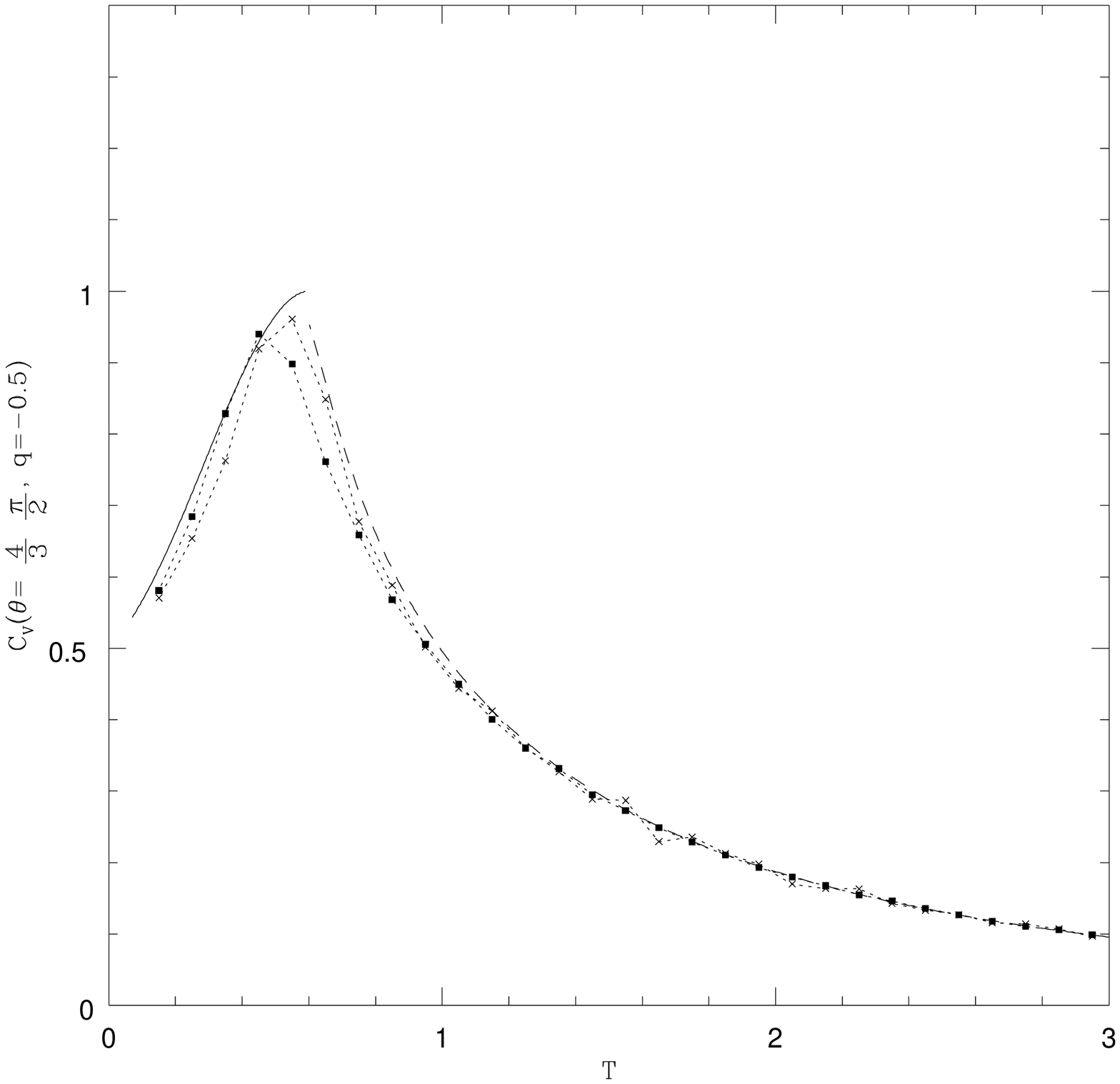}
\protect\caption[1]{
As in fig.  (\protect\ref{F_CVQM233ALL}), but $q=-.5$.
\protect\label{F_CVQM5ALL}}
\end{figure}

We show in figures (\ref{F_EQM5ALL}) and (\ref{F_CVQM5ALL}) that even at
very high negative values of $q$ (i.e. at least down to $q=-0.5$) our replica
solution of the model with quenched disorder gives a very accurate
description of the behavior of the deterministic model with complex frustration
in the low $T$ phase. Even the specific heat very close to the critical point
is reconstructed with good accuracy.

\begin{figure}
\epsfxsize=400pt\epsffile{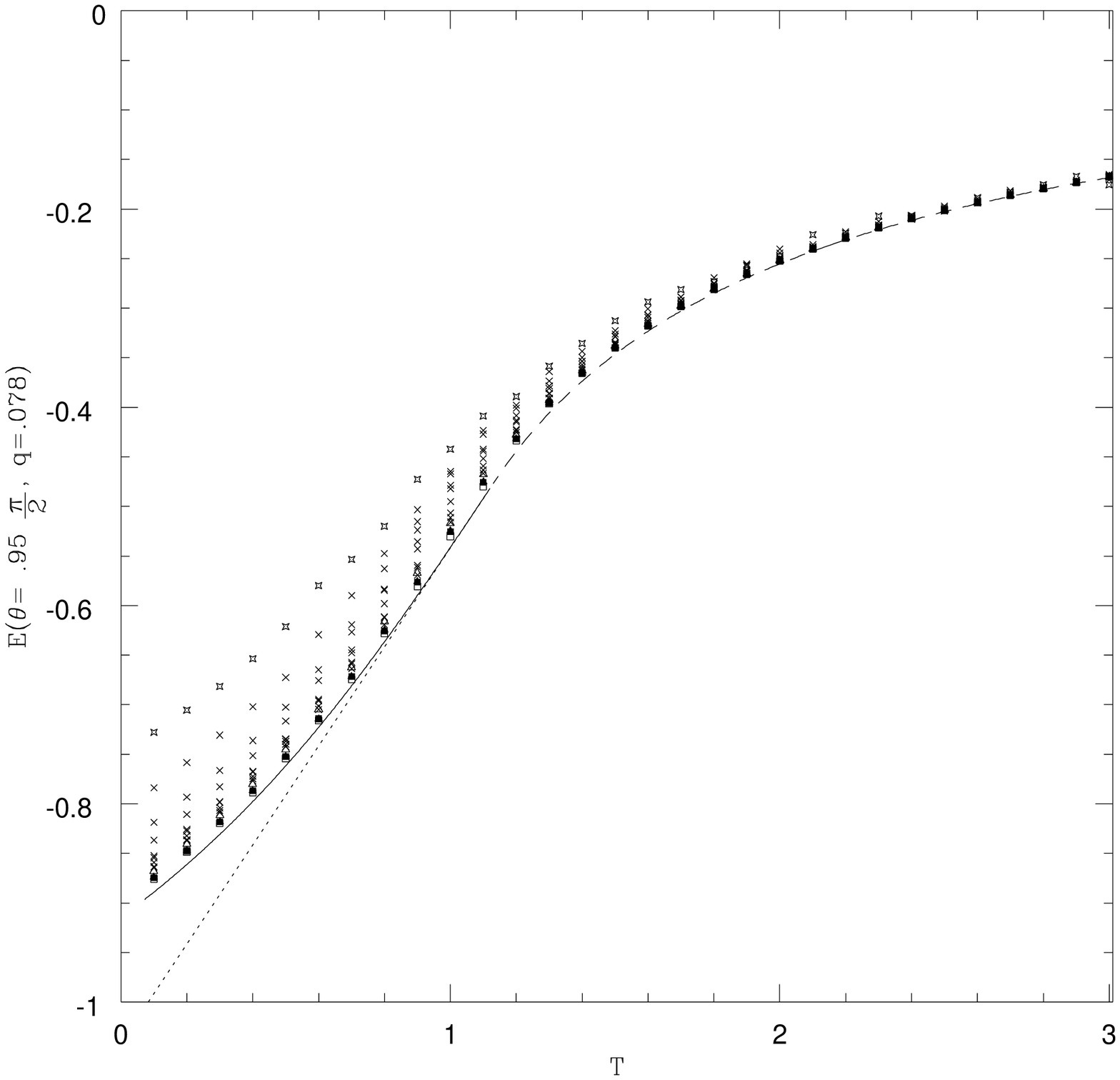}
\protect\caption[1]{
As in fig.  (\protect\ref{F_EQ0ALL}), but for $q=.078$. Here the dashed
straight line is the result one would get for the spherical model, where the
energy becomes linear in $T$ below the critical point.
\protect\label{F_EQP78ALL}}
\end{figure}

\begin{figure}
\epsfxsize=400pt\epsffile{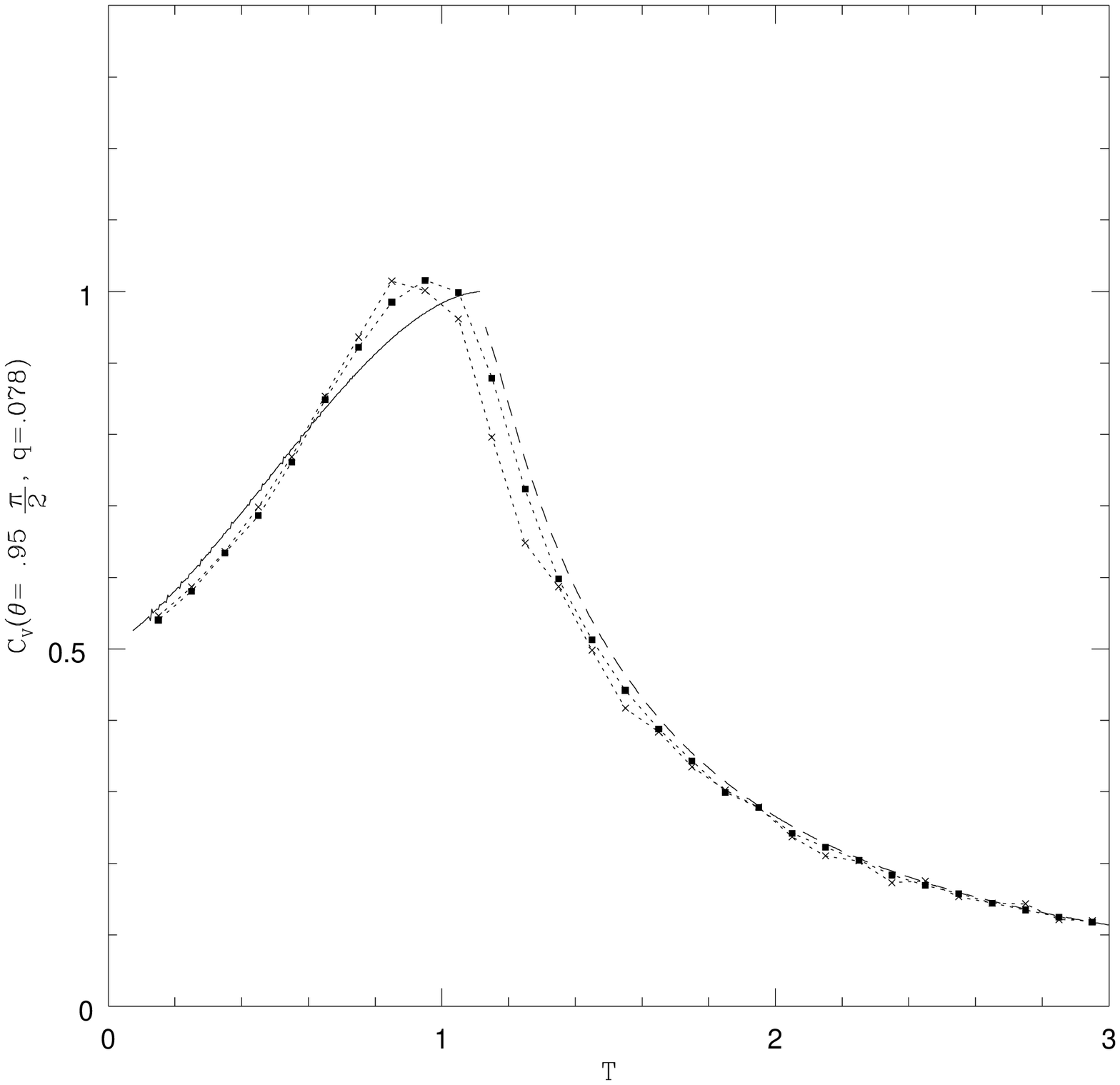}
\protect\caption[1]{
As in fig.  (\protect\ref{F_CVQM233ALL}), but $q=.078$.
\protect\label{F_CVQP78ALL}}
\end{figure}

The situation is different on the side of $\gth<\frac{\pi}{2}$, i.e. for
positive values of $q$. At low positive $q$ there are again no dramatic
problems, and if the two models differ they do differ only in a very minor
way. In fig.  (\ref{F_EQP78ALL}) we add a dashed straight line, from $T_c$
down to $T=0$, to give the result one would obtain for the spherical model
\cite{SPHERICAL}, where the energy becomes linear in $T$ below the critical
point. In fig. (\ref{F_CVQP78ALL}) we plot again the specific heat.  If there
is a discrepancy it small, even if we want already to notice the small bump
just under $T_c$, which makes the Monte Carlo data slightly different from the
disordered model result. This effect was not there for negative $q$ values,
and it is not clear here if it is due to a true difference or if it is
connected to a finite size effect.

\begin{figure}
\epsfxsize=400pt\epsffile{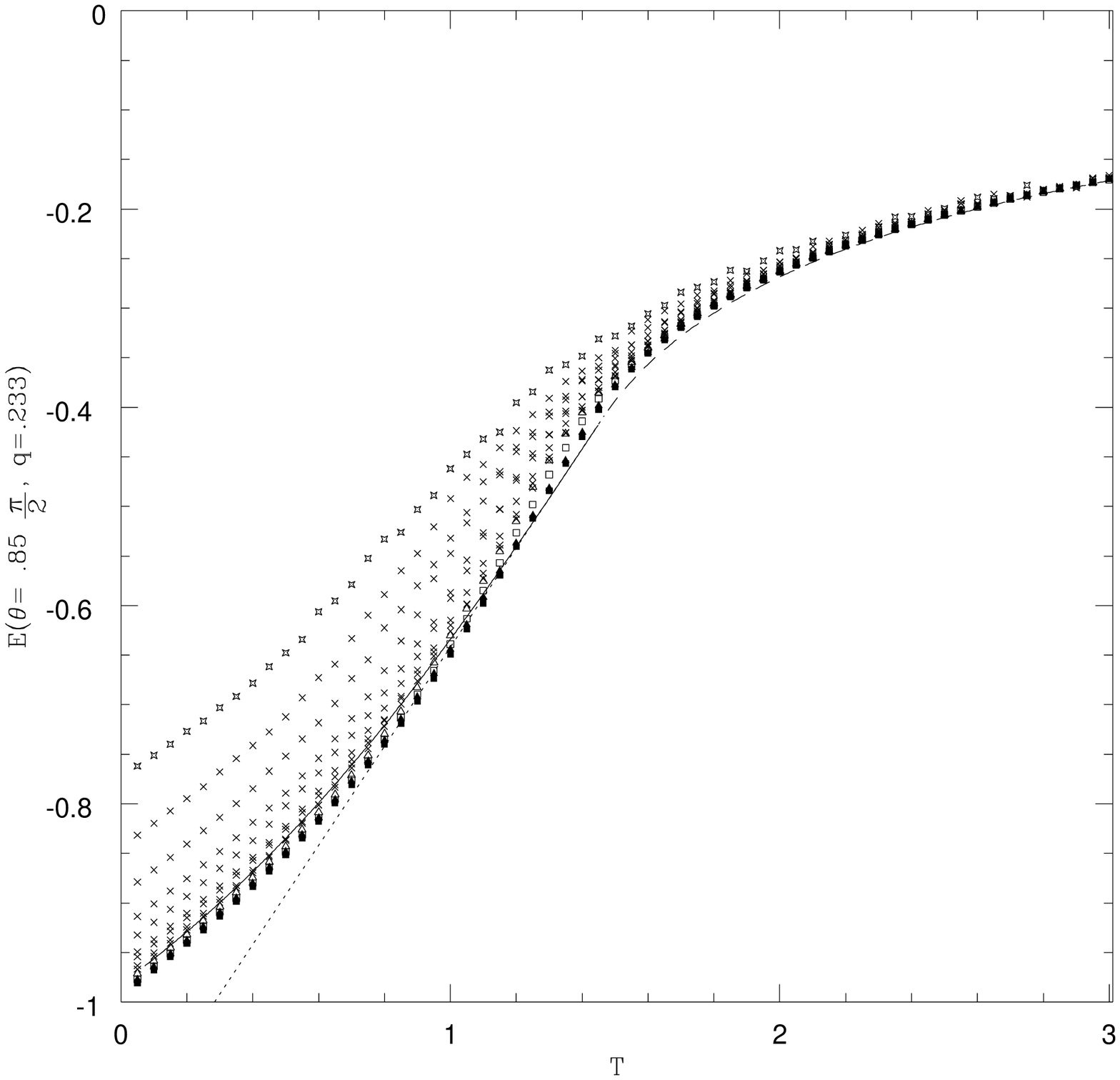}
\protect\caption[1]{
As in fig.  (\protect\ref{F_EQ0ALL}), but for $q=.233$.
\protect\label{F_EQP233ALL}}
\end{figure}

\begin{figure}
\epsfxsize=400pt\epsffile{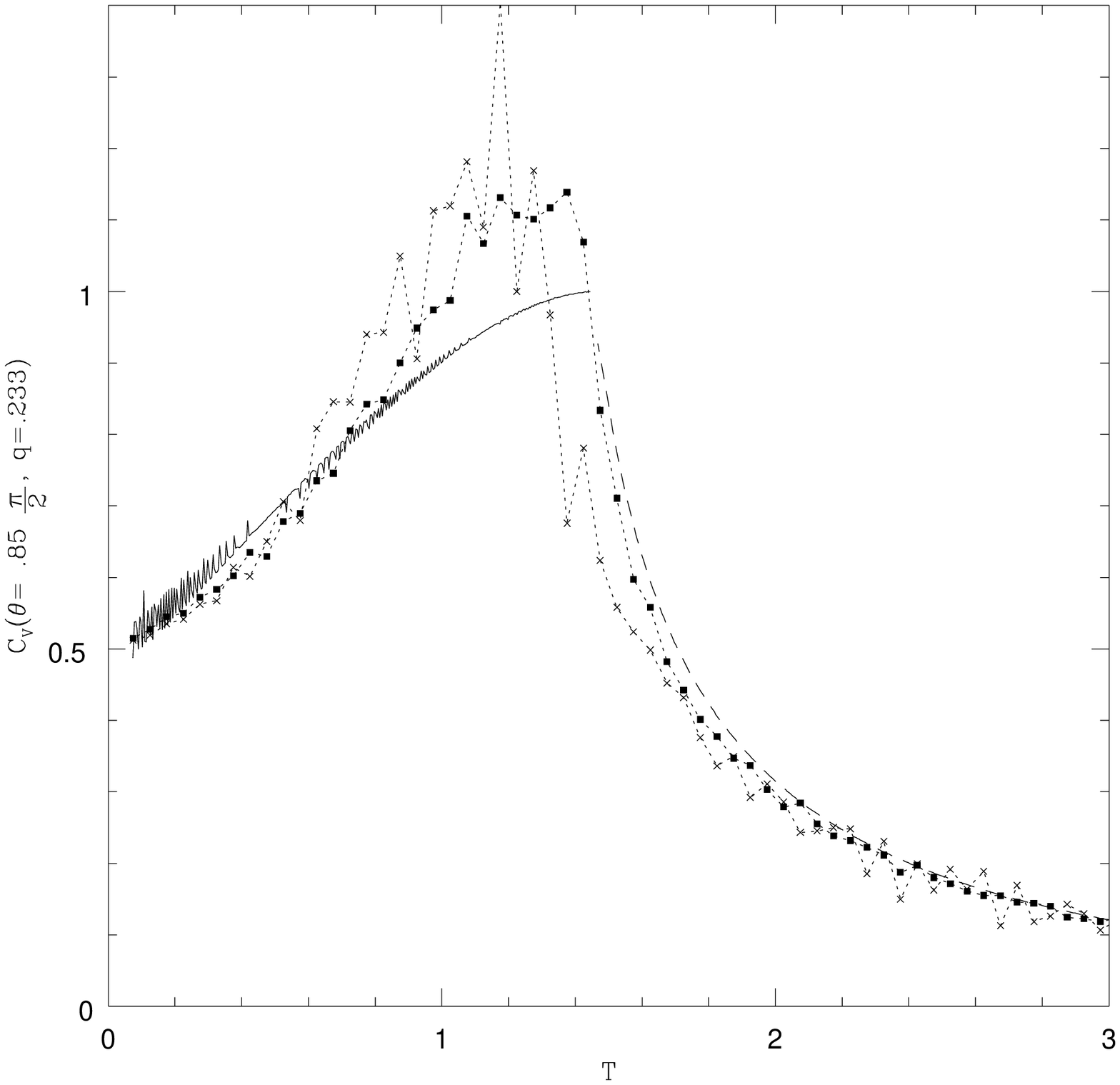}
\protect\caption[1]{
As in fig.  (\protect\ref{F_CVQM233ALL}), but $q=.233$.
\protect\label{F_CVQP233ALL}}
\end{figure}

The situation becomes more clear (in a negative sense) when we increase $q$
of a not huge amount. We give in figures (\ref{F_EQP233ALL}) and
(\ref{F_CVQP233ALL}) the results for $q=.233$, and here there is a clear
discrepancy, which is difficult to justify by means of finite size effects.
Indeed here the energy of the Monte Carlo simulations at low $T$ is, already
for $D=16$ lower than the analytic result one gets for the spherical
random model in the infinite volume limit. Since the energy is decreasing with
$D$, and we expect the energy of the spherical model to be a lower bound at
all $T$ to our XY case, that seems to show that in this case the two models
do indeed differ, even if only of a small amount. In order to explain this
effect one would have to assume that the sign of the corrections changes with
the dimensionality, and that the energy will go up again for $D$ large enough.
This is not impossible, but not so plausible, and we have no numerical
indications for such an effect to be taking place. The specific heat picture
(\ref{F_CVQP233ALL}) is even more self-explanatory than the energy, since it
is quite difficult to believe that the big bump of the Monte Carlo data will
be reabsorbed in the $D\to\infty$ limit.

In conclusion, it seems that for $q<0$ and even for small $q$ positive
values the replica theory describes the deterministic model with very high
accuracy. On the contrary for $q>0$ not so small there is a clear, even if
quite small discrepancy between the two models.

\section*{Appendix}
In this short appendix we will fill a gap in the proof of equation
(\ref{MAGIC}). We only sketch the main steps of the proof, which is absolutely
inelegant. It is quite likely that a more elegant proof, e.g. based on the
braid group, do exist, but we have not found it.

In \cite{PARISI} it was proved that

\be
  G_k^{(q)} \equiv \int d\gl\ \gr_\gD(\gl)\ \gl^n
  = \sum_{n=0,\infty}\cN(k,n) q^n\ ,
\ee
where $\cN(k,n)$ is the number of ways in which one can connect
piecewise $k$ points on a circle, with $n$ intersections.

In order to compute $\cN(k,n)$ it may convenient to consider the quantity
$\cN(k,n,m)$, i.e. the number of ways in which $k+1$ points on the circle may
be connected in such a way that a line starts from each of the first $k$
points and $m$ lines arrive in the last $k+1$-th point, the total number of
intersection being $n$. It is evident that

\be
\cN(k,n)=\cN(k,n,0)\ .
\ee
A simple pictorial argument can be used to prove that

\be
\cN(k+1,n,m)= \cN(k,n,m-1)+ \sum_{j=0,m}\cN(k,n-j,m+1)\ .
\ee
We can now check that this relation is satisfied if we set

\be
  \sum_{n=0,\infty}\cN(k,n,m) q^n \equiv G_k^{(q)}(m)=\lan m|Y^k|0\ran
\ee
where the $| n\ran$ (for $n=0,\infty$) form a basis in an Hilbert space,

\be
  Y \equiv A+A^\dagger \ ,
\ee
and

\bea
  A|n\ran &=& |n-1\ran \ \mbox{for} \ n\NE 0\ , \\
  A|0\ran &=& 0 \ ,\\
  A^\dagger|n\ran &=& \frac{1-q^{n+1}}{1-q} |n+1 \ran \ . \nonumber
\eea
The operator $X$ in eq. (\ref{MAGIC}) and $Y$ are related by the simple
transformation $X=MYM^{-1}$, where the operator $M$ is diagonal in the
basis we have used. We finally find that

\be
  G_k^{(q)}(0) = \lan m|Y^k|0\ran =\lan m|X^k|0\ran\ ,
\ee
which is the result announced in \cite{PARISI}.

\end{document}